\documentclass[a4paper,11pt]{article}
\usepackage{graphicx} % Required for inserting images

\pdfoutput=1 % if your are submitting a pdflatex (i.e. if you have
\usepackage{jcappub} % for details on the use of the package, please
\usepackage[T1]{fontenc} % if needed

\usepackage[utf8]{inputenc}
\usepackage{amsmath,mathtools}

\usepackage{tensor}
\usepackage{amssymb}
\usepackage{graphicx}
\usepackage{physics}
\usepackage{lipsum}  
\usepackage{bigints}
\usepackage{verbatim}

\usepackage{braket}
\usepackage{cancel}
\usepackage{booktabs}
\usepackage{color}
\usepackage{hyperref}
\usepackage{graphicx,subcaption}

\newcommand{\roughly}[1]{\mathrel{\raise.3ex\hbox{$#1$\kern-0.85em
\lower1ex\hbox{$\sim$}}}}
\newcommand{\lsim}{\roughly<}
\newcommand{\gsim}{\roughly>}

\newcommand{\mfa}{{\mathfrak a}}
\newcommand{\mfm}{{\mathfrak m}}
\newcommand{\bfa}{{\mathbf{a}}}
\newcommand{\bfg}{{\mathbf{g}}}
\newcommand{\cG}{{\cal G}}

\newcommand{\cL}{{\cal L}}
\newcommand{\cO}{{\cal O}}

\newcommand{\ssB}{{\scriptscriptstyle B}}

\newcommand{\ssQ}{{\scriptscriptstyle Q}}
\newcommand{\ssR}{{\scriptscriptstyle R}}
\newcommand{\ssT}{{\scriptscriptstyle T}}

\newcommand{\MPL}{M_{p}}
\newcommand{\MEW}{M_{\scriptscriptstyle W}}
\newcommand{\exd}{{\rm d}}
\newcommand{\pref}[1]{{(\ref{#1})}}

\newcommand{\B}{{\ssB}}

\newcommand{\s}{{\chi}}
\newcommand{\ax}{{\mfa}}
\newcommand{\sax}{{\rm ax}}
\newcommand{\bax}{{\mathbf{a}}}
\newcommand{\axm}{{\mfm}}
\newcommand{\nn}{\nonumber}

\begin{document}

\title{A Minimal Axio-dilaton Dark Sector}

\author[a]{Adam Smith,}
\affiliation[a]{School of Mathematical and Physical Sciences, University of Sheffield%, Hounsfield Road, Sheffield S3 7RH, United Kingdom
}
\author[b,c]{Maria Mylova,}
\affiliation[b]{Kavli Institute for the Physics and Mathematics of the Universe (WPI)%, The University of Tokyo, Kashiwa, Chiba 277-8583, Japan
}
% \affiliation[c]{Science Education, Ewha Womans University%, 52 Ewhayeodae-gil, Seoul, Republic of Korea
% }
\affiliation[c]{Perimeter Institute for Theoretical Physics%, 31 Caroline Street North, Waterloo ON, Canada.
}
\author[d]{Philippe Brax,}
\affiliation[d]{Institut de Physique Th\'eorique, Universit\'e Paris-Saclay%,
%CEA, CNRS, F-91191 Gif-sur-Yvette Cedex, France.
}

\author[a]{Carsten van de Bruck,}

\author[c,e,f]{C.P.~Burgess}

\affiliation[e]{Department of Physics \& Astronomy, McMaster University%, 1280 Main Street West, Hamilton ON, Canada.
}
\affiliation[f]{School of Theoretical Physics, Dublin Institute for Advanced Studies%, 10 Burlington Rd., Dublin,  Co. Dublin, Ireland
}
\author[g,h]{and Anne-Christine Davis}
\affiliation[g]{DAMTP, University of Cambridge%, Wilberforce Road,  Cambridge, CB3 0WA, UK.
}
\affiliation[h]{Kavli Institute of Cosmology (KICC), University of Cambridge%, Madingley Road, Cambridge, CB3 0HA, UK. 
}

\emailAdd{asmith69@sheffield.ac.uk}
\emailAdd{mylova@g.ecc.u-tokyo.ac.jp}
\emailAdd{philippe.brax@ipht.fr}
\emailAdd{c.vandebruck@sheffield.ac.uk}
\emailAdd{cburgess@perimeterinstitute.ca}
\emailAdd{ad107@cam.ac.uk}

\abstract{
  In scalar-tensor theories it is the two-derivative sigma-model interactions that like to compete at  low energies with the two-derivative interactions of General Relativity (GR) -- at least once the dangerous zero-derivative terms of the scalar potential are suppressed (such as by a shift symmetry). But nontrivial two-derivative interactions require at least two scalars to exist and so never arise in the single-scalar models most commonly explored. Axio-dilaton models provide a well-motivated minimal class of models for which these self-interactions can be explored. We review this class of models and investigate whether these minimal two fields can suffice to describe both Dark Matter and Dark Energy. We find that they can -- the axion is the Dark Matter and the dilaton is the Dark Energy -- and that they robustly predict several new phenomena for the CMB and structure formation that can be sought in observations. These include specific types of Dark Energy evolution and small space- and time-dependent changes to particle masses post-recombination that alter the Integrated Sachs-Wolfe effect, cause small changes to structure growth and more. 
}

\maketitle
\section{Introduction}

Scalar-tensor models have been used for decades \cite{Jordan:1955, Brans:1961sx, Dicke:1961gz, Brans:2014} as foils for Einstein's General Relativity (GR) when testing our understanding of gravity observationally. They provide plausible modifications to GR whose predictions can differ in interesting ways and using observations to constrain scalar-tensor couplings can quantify the accuracy with which we believe GR to be true  (for reviews see \cite{Will:2014kxa, Berti:2015itd}).

\subsection{The case for axio-dilatons}

The vast majority of these studies (at least for tests of gravity in the solar system and in the late universe) quite sensibly take a minimalist approach and modify gravity by adding only a single real scalar field (call it $\phi$), see e.g. \cite{Wetterich:1987fm,Damour:1990tw,Damour:1993id,Wetterich:1994bg,Gasperini:2001pc,Khoury:2003rn, Brax:2004qh,Hinterbichler:2010es,Brax:2010gi,Brax:2011ja} and see \cite{Fujii:2003pa,Burrage:2017qrf} for a review and further references. In this case the lagrangian can be cast as a derivative expansion for which the most general local form is\footnote{Any function of $\phi$ appearing in front of the Ricci scalar can be removed by performing an appropriate field redefinition of the metric and the scalar kinetic term is put into a standard form by redefining $\phi$. }
\begin{equation} \label{SingleFieldL}
    \cL = - \sqrt{-g} \Bigl[  V(\phi) + \tfrac12 \, \MPL^2 R + \tfrac12 \, \MPL^2 \partial_\mu \phi \, \partial^\mu \phi + \cdots \Bigr] + \cL_m(\psi, \phi, g_{\mu\nu}) \,,
\end{equation}
where $\cL_m$ contains all terms involving ordinary Standard Model fields -- denoted collectively here by $\psi$ -- while $V(\phi)$ is the scalar potential, $R$ the spacetime metric's Ricci scalar and the ellipses represent higher-derivative terms. Our conventions choose $M_p^{-2}=8\pi G$ where $G$ is Newton's constant of universal gravitation.

Of these, it is known that the most important interactions at low energies come from $V(\phi)$ and these are so important that they are dangerous. They can easily mediate interactions much stronger than gravity and some sort of approximate symmetry is usually assumed in order to suppress their otherwise too-large contributions. Naturalness issues associated with explaining why the scalar can be light enough to play a role in tests of gravity are a particular aspect of this story. 

Such arguments can be systematized as power-counting estimates within effective field theories (EFTs) and when applied to scalar-tensor theories \cite{Burgess:2009ea, Adshead:2017srh} these tell us several things. First and foremost these arguments explain why the classical approximation often works so well in gravity: the semiclassical approximation (loop expansion) in gravity is at heart a low-energy derivative expansion. Organizing the lagrangian as in \pref{SingleFieldL} is not only convenient; it is often {\it compulsory}. When the derivative expansion for $\cL$ breaks down so do the classical methods we use to extract its predictions (there can be exceptions -- see for instance \cite{Babic:2019ify}). 

For single-field models we learn from \pref{SingleFieldL} that if the scalar potential is for some reason not present (perhaps due to a symmetry under constant shifts of $\phi$) then the minimally coupled gravitational interactions are the only ones possible that involve only $\phi$ and the metric and only have two derivatives. All other scalar-metric interactions involve higher derivatives and are therefore suppressed at the low energies where tests of gravity in practice take place.

This same argument is {\it not} true for multi-scalar scalar-tensor models because once more than one field, $\phi^a$, is present the most general derivative expansion for $\cL$ replaces \pref{SingleFieldL} with
\begin{equation} \label{MultiFieldL}
    \cL = -\sqrt{-g} \Bigl[  V(\phi) + \tfrac12 \, M_p^2 R + \tfrac12\, M_p^2 \, \cG_{ab}(\phi) \, \partial_\mu \phi^a \, \partial^\mu \phi^b + \cdots \Bigr] + \cL_m(\psi, \phi, g_{\mu\nu}) \,,
\end{equation}
where the function $\cG_{ab}(\phi)$ introduces the possibility of nontrivial new two-derivative interactions. These interactions are special because they are not suppressed relative to the two-derivative interactions of GR even at low energies. This makes multi-scalar models particularly interesting when interpreting low-energy tests of GR and two-scalar models are perhaps of most interest as their simplest possible representatives.  

Axio-dilaton models are a particularly well-motivated class of minimal two-scalar theories for which the interactions mediated by $\cG_{ab}(\phi)$ are possible. For the present purposes these can be defined as models involving two real scalars, $\chi$ and $\mfa$, for which the\footnote{We break with the convention that reserves the word `axion' specifically for the QCD axion \cite{Weinberg:1977ma, Wilczek:1977pj} that arises within the context of the Strong-CP problem \cite{Peccei:1977hh} and use it indiscriminately for any axion-like particle (ALP).} `axion' $\mfa$ is a pseudo-Goldstone boson for an internal symmetry under which $\mfa$ shifts by a constant so $\cG_{ab}(\chi)$ is $\mfa$-independent. The most general possible two-derivative scalar self-interaction consistent of this type has the form \pref{MultiFieldL} with 
\begin{equation} \label{SL2RKinTerm}
    \cG_{ab}(\phi) \, \partial_\mu \phi^a \, \partial^\mu \phi^b = \partial_\mu \chi \, \partial^\mu \chi + W^2(\chi) \, \partial_\mu \mfa \, \partial^\mu \mfa ~.
\end{equation}
The `dilaton' $\chi$ is also a pseudo-Goldstone boson, but for an approximate spacetime scaling symmetry. Under approximate scaling symmetries the classical lagrangian scales by a constant factor when the metric is scaled by a factor and $\chi$ is shifted by a constant amount (and $\mfa$ is scaled appropriately). Although the action is {\it not} invariant under such a transformation, the classical equations of motion typically are (and this is why such symmetries are ultimately useful \cite{Burgess:2020qsc}).

These scaling conditions typically lead $W(\chi)$ and any scalar potential $V(\chi)$ to be exponential in form, 
\begin{equation} \label{ExponentialChoices}
    W(\chi) = W_0 \, e^{-\zeta \chi} \quad \hbox{and} \quad
    V_{\rm dil}(\chi) = U \, e^{- \lambda \chi} \,,
\end{equation}
which is a choice we also make here (though we first develop our field equations in the general case). It is always possible to shift $\chi$ to ensure $W_0 = 1$ (and more about the prefactor $U$ below).

In practice the scaling symmetry also usually means $\chi$ couples to matter only through a Jordan-frame metric
\begin{equation} \label{JFtoEF}
    \cL_m(\Psi, \chi, \mfa, g_{\mu\nu}) = \cL_m(\Psi, \partial_\lambda\mfa, \tilde g_{\mu\nu}) \quad \hbox{where} \quad
    \tilde g_{\mu\nu} := C^2(\chi) \, g_{\mu\nu} \,,
\end{equation}
where $C(\chi) = e^{\bfg \chi}$ is also exponential in $\chi$. Once combined with eqs.~\pref{SL2RKinTerm} and \pref{MultiFieldL} this means that $\chi$ couples to matter in the same way as does a Brans-Dicke scalar \cite{Brans:1961sx, Dicke:1961gz}, with Brans-Dicke parameter $\omega$ related to $\bfg$ by $2\bfg^2 = (3 + 2 \omega)^{-1}$.  

This combination of scalar fields arises very commonly amongst the low-energy fields in string vacua and this is true for robust symmetry reasons: the approximate symmetries for which they are the low-energy pseudo-Goldstone bosons are generic in string compactifications \cite{Burgess:2020qsc}. In supersymmetric models axio-dilatons generically appear together within the same supermultiplet (the simplest of which must have at least two scalars).

Supersymmetry can also predict relationships amongst the parameters $\bfg$ and $\zeta$, and we give examples of this in \S\ref{sec:DilatonProperties} below. In particular for a broad class of well-motivated models, scale invariance together with supersymmetry predict $\bfg = - \frac12 \alpha \zeta = - \sqrt{\frac{\alpha}{6}}$ where $\alpha$ is a real parameter. Yoga models \cite{Burgess:2021obw} are a specific subset of this class that also predicts $\lambda = 4 \zeta$ and $\alpha = 1$. In what follows we use the relation $\lambda = 4 \zeta$ and allow $\bfg$ and $\zeta$ to be independent phenomenological parameters. 

One might ask how supersymmetry could possibly be relevant at the extremely low energies relevant to cosmology given the absence of evidence for it at the much higher energies pertinent to particle physics. This can easily arise within supersymmetric theories because the splitting of masses between bosons and fermions within a supermultiplet takes the form
\begin{equation}
    \Delta m^2 \sim g F
\end{equation}
where $F$ is the supersymmetry-breaking expectation value and $g$ is the coupling strength between $F$ and the multiplet whose splittings are being calculated. Since gravity is the weakest interaction, gravitationally coupled supermultiplets often get split by much less than the other, more strongly coupled, supermultiplets. It would not be at all unusual to have a gravitationally coupled dark sector that is approximately supersymmetric at energies well below particle physics scales (a possibility that is further explored in \cite{Burgess:2021juk}).  

\subsection{The Dark Axio-dilaton }

To this point we have argued that a minimum of two fields are required in order for scalar-tensor models to have two-derivative interactions that can compete in an interesting way with the two-derivative interactions of GR. Axio-dilatons are a well-motivated minimal example, and so are particularly interesting as alternatives to gravity when testing GR at low energies. In this paper we make the case that this minimal combination could also suffice to provide a compelling minimal model {\it for the entire observed Dark Sector}\footnote{A similar motivation and idea has been discussed in \cite{Bernardo:2022ztc}.}. 

By and large GR emerges victorious whenever its predictions are tested, with the two main exceptions being the suites of observations that point to the existence of Dark Matter and Dark Energy. These not only show that there is something out there gravitating that we do not understand; they show there are {\it two} types of unknown things doing so. Could the two scalars needed to get interesting low-energy competition with gravity also be the two mysterious gravitating quantities in cosmology? If so they would provide a compellingly minimal modification to GR that might capture what we see around us. 

Part of what makes this proposal attractive is the observation that the axion and dilaton have separately already been suggested as a Dark Matter or Dark Energy candidate. On one hand, axions have long been a leading candidate for Dark Matter \cite{Preskill:1982cy, Abbott:1982af, Dine:1982ah} (for a review see \cite{Marsh:2015xka}) because coherent field oscillations have an energy that falls within an expanding universe in the same way as does a nonrelativistic species of particle. The attractiveness of axions as dark-matter candidates has only grown as the viability of WIMP Dark Matter has decreased \cite{Arcadi:2024ukq}.

On the other hand a Brans-Dicke scalar like the dilaton of an axio-dilaton pair is a favourite amongst quintessence models for Dark Energy (for a review see \cite{Copeland:2006wr}) because they tend to generate scalar potentials that are exponential in $\chi$ and these can provide good descriptions for Dark Energy if they are not too steep \cite{Ratra:1987rm}. A common objection to quintessence models is that they have naturalness problems (not least of which is the cosmological constant problem). But dilatons also play a central role in models that use scale invariance to try to address these issues, such as in Yoga models \cite{Burgess:2021obw} that exploit the nontrivial low-energy interplay between supersymmetry and scaling symmetries to try to construct a natural relaxation mechanism. 

The precise question we here ask is: can the axio-dilaton provide the one-stop shopping cosmologists need to describe both Dark Matter and Dark Energy? More specifically, recent studies show \cite{Smith:2024ayu} that axio-dilaton models can produce acceptable cosmologies provided that Dark Matter is already present as a separate ingredient. We here ask whether the axion part of the axio-dilaton already used to describe Dark Energy can itself provide a minimal and predictive description of Dark Matter. To answer this question we build on the previous work in \cite{Smith:2024ayu}, where the axion in all cases but one was taken to be massless, by giving the axion part an appropriate quadratic scalar potential, as known to be compatible with scalar field dark matter models \cite{Marsh:2015xka}.

Our answer to this question is: `yes', at least so far as cosmology is concerned. We argue this by showing that the energy density of minimal axio-dilaton cosmologies can have the properties needed to describe Dark Matter without losing the successful dilaton description of Dark Energy. Moreover they can do so using parameters that are known to be consistent with solar-system tests of GR \cite{Will:2014kxa, Berti:2015itd}. 

To this end \S\ref{SecII:MinimalAxioDilaton} specifies more precisely the properties of these models that we need. This starts in \S\ref{sec:CosmicEvolution} which evaluates the evolution equations describing cosmological evolution, for both backgrounds and the linearized fluctuations around them. This is done using completely general choices for the scalar potential and the kinetic function $W(\chi)$. 

The necessity of rapid axion oscillations is an important complication for numerical evolution so we handle these using the standard Madelung framework \cite{Madelung:1927ksh} that treats the rapid oscillations as a fluid in the presence of the averaged scalar. The application of this formalism to axio-dilatons is spelled out in detail in \S\ref{SecIII:MinimalDarkSector}. Our choices for the dilaton are then given in \S\ref{sec:DilatonProperties}, and those made for the axion potential and matter couplings are given in \S\ref{sec:AxionProperties}. The absence of direct axion-matter couplings makes the axion a run-of-the mill ALP from the point of view of its couplings to gravity, whose energy density in cosmology is dominated by its small oscillations about a local minimum of its potential (for the distinction amongst, and constraints on, ALP models see \cite{ParticleDataGroup:2024cfk}). As we shall see, the two-derivative interactions generically cause the Dark Matter and Dark Energy to interact at low energies even in the absence of a scalar potential. 

\S\ref{SecIV:Numerics} then verifies that there exist background solutions that provide broadly acceptable cosmologies inasmuch as they properly capture the observed evolution of the known cosmological fluids, with the axion energy density interpreted as Dark Matter and the dilaton energy density as Dark Energy. The fluctuation equations are then implemented into a modification of the Cosmic Linear Anisotropy Solving System code (CLASS) \cite{Diego_Blas_2011}, which we use to compute both the matter power spectrum and the CMB angular spectra.  

Plots for cosmologies are provided in figures \ref{fig:No well} and \ref{fig:Well} for a variety of choices for the axion-dilaton coupling $\zeta$ assuming the dilaton-matter coupling is $\bfg = -10^{-3}$ (and so is small enough to satisfy solar system constraints on Brans Dicke couplings \cite{Will:2014kxa, Bertotti:2003rm}). The resulting plots disagree with $\Lambda$CDM predictions for sufficiently large $\zeta$ but can become acceptable by eye once $\zeta$ is as small as 0.01. We do not perform a fit of our predictions to CMB and structure formation data, but this would be an interesting next step in this analysis.

Most interestingly, axiodilaton models provide several robust predictions that can be sought to help determine if this is the way Nature really works. Because the dilaton behaves as a quintessence field it generically implies a time-varying Dark Energy equation of state. The two-derivative interactions provide a robust DM-DE interaction that scales at low energies in the same was that GR does.

Another class of predictions follows from \pref{JFtoEF}, which implies that all Standard Model particles acquire $\chi$-dependent masses in Einstein frame, with $m(\chi) = m_0 \, e^{\bfg \chi}$. We show how the scale invariance of the setup implies that $\chi$ likes to evolve in `tracker' power-law solutions \cite{Ferreira:1997hj}, whose nature inevitably changes as the universe passes from radiation to matter domination. When it does so $\chi$ inevitably makes a transient excursion that is still taking place at recombination. 

This implies a generic prediction of small universal variations in particle masses (in Planck units) between now and recombination whose size is controlled by $\bfg$. For $|\bfg| = 10^{-3}$ the predicted mass change is slightly less than a part per mille in the CMB, with masses slightly larger at recombination than they are now. Whether there is a similar effect at nucleosynthesis depends somewhat on the form of the dilaton's scalar potential. Because this excursion is not completed before recombination it can produce observable effects like a modified ISW effect that disfavours large values of $\zeta$. We discuss in \S\ref{sec:DilatonProperties} why position- and time-dependent mass effects are predicted to be much smaller on Earth, escaping present-day constraints.

\S\ref{SecIV:Numerics} closes with the calculation of potentially observable effects for the growth of structure. Deviations from $\Lambda$CDM arise due to the existence of dilaton exchange providing a new attractive channel and because axio-dilaton interactions cause the dilaton excursion to feed back into the axion ({\it i.e.}~Dark Matter) mass. We compute the implications for an observable related to $\sigma_8$ and show that it can be larger than or smaller than the $\Lambda$CDM prediction at present, but in a way that correlates with the changes in particle masses.

Our conclusions are summarized in \S\ref{sec:Conclusions}.

\section{Axio-dilaton models}
\label{SecII:MinimalAxioDilaton}

We start with a brief reminder of the field equations for two-scalar models whose cosmological solutions are of interest (see \cite{Smith:2024ayu} for more details). Consider first a general two-scalar model for which we assume only that the target-space metric is independent of one of the fields $\cG_{ab}(\chi,\mfa) = \cG_{ab}(\chi)$ (as can be enforced using a shift symmetry) and that ordinary matter couples only to $\chi$ only through a Jordan-frame metric, $\tilde g_{\mu\nu}$, that is related to the Einstein-frame metric $g_{\mu\nu}$ as in \pref{JFtoEF}. 

The action for such a model is
\begin{equation}\label{Action}
    S=- \int d^4x\sqrt{-g}\Bigl\{ \tfrac{1}{2} M_p^2 \Bigl[ R + \partial_\mu \s \, \partial^\mu\s + W^2(\chi) \, \partial_\mu\ax \, \partial^\mu\ax \Bigr] + V(\chi,\ax)\Bigr\} + S_m(\Psi, \tilde g) \,.
\end{equation}
and so the Einstein Equations for such a system are given by
\begin{align}
\label{einstein eq}
 & G_{\mu\nu}-\left(\partial_\mu \chi \, \partial_\nu \chi - \frac{1}{2} g_{\mu\nu} \, \partial^\sigma \chi  \,\partial_\sigma \chi \right) \nonumber
\\& \qquad\qquad\qquad -
W^2 \left( \partial_\mu \mfa  \,\partial_\nu \mfa - \frac{1}{2} g_{\mu\nu} \, \partial^\sigma \mfa \, \partial_\sigma \mfa \right) 
+\frac{1}{\MPL ^2} \Bigl( V\, g_{\mu\nu}  - T_{\mu\nu} \Bigr) = 0 \, ,  
\end{align}
where $G_{\mu\nu}$ is the metric's Einstein tensor, $T^{\mu\nu} = 2 (-g)^{-1/2} (\delta S_m/\delta g_{\mu\nu})$ denotes the matter sector's Einstein-frame stress-energy tensor and $T = g_{\mu\nu} T^{\mu\nu}$ is its trace. The field equations for the dilaton and axion fields similarly are
\begin{equation}
\label{dilaton eom}
    \Box\chi -W \, W,_\chi \partial_\mu\mfa \, \partial^\mu\mfa - \frac{V,_\chi}{\MPL ^2} = -\frac{\mathbf{g}T}{\MPL ^2}  \, ,
\end{equation}
and
\begin{equation}
\label{axion eom}
 \Box\mfa + \frac{2W,_\chi}{W} \, \partial_\mu\chi \, \partial^\mu\mfa - \frac{V,_\mfa}{W^2\MPL ^2} = 0 \,.
\end{equation}
Subscripts $_,\chi$ and ${_,\mfa}$ respectively denote differentiation with respect to the corresponding fields. 

The model is fixed by specifying the functions $W(\chi)$, $V(\chi,\mfa)$ and the size of the dilaton-matter coupling $\bfg$. For cosmological applications the stress energy can be chosen to be the sum of contributions with the perfect-fluid form
\begin{equation}
     T_{(f)}^{\mu\nu}   = (\rho_f + p_f) u^\mu_f u^\nu_f + p_f g^{\mu\nu} \,,
\end{equation}
where $f = B,R$ for baryon and radiation fluids while $\rho_f$ and $p_f$ are the corresponding energy density and pressure and $u^\mu_f$ is the fluid's local 4-velocity. 

\subsection{Cosmic evolution}
\label{sec:CosmicEvolution}

We next obtain the evolution equations governing late-time cosmology. Since these equations are derived in detail in \cite{Smith:2024ayu} we skip many details and simply quote the main results here.

\subsubsection*{Background dynamics}

Assuming a homogeneous and isotropic background allows us to specialize to the metric
\begin{equation}
    {\rm d}s^2 = -{\rm d}t^2 + a^2\,  \delta_{ij} \, {\rm d}x^i {\rm d}x^j  
    = a^2\left[-{\rm d}\eta^2 + \delta_{ij} \, {\rm d}x^i {\rm d}x^j\right] \, ,
\end{equation}
where we take flat spatial slices and  $a=a(\eta)$ is a function of cosmic time $t$ or conformal time $\eta$ (related as usual by $\exd t = a \, \exd \eta$). Background scalar fields are also taken to depend only on time, $\chi = \bar \chi(\eta)$ and $\mfa = \bar\mfa(\eta)$.

With these assumptions the field equations for the background reduce to the following ordinary differential equations
\begin{equation} \label{friedman bg}
    \mathcal{H}^2 = \frac{1}{3\MPL ^2}\left[\left(\frac{\Bar{\chi}'^2}{2}+\frac{W^2\Bar{\mfa}'^2}{2}\right)\MPL ^2+a^2V+a^2\Bar{\rho}\right] \, ,
\end{equation}
\begin{equation}  \label{friedmann dilaton}
     \Bar{\chi}''+2\mathcal{H}\Bar{\chi}'-WW,_\chi\Bar{\mfa}'^2+\frac{a^2}{\MPL ^2}\Bigl( V,_{\chi} +\bfg  \, \Bar{\rho}_\ssB  \Bigr)=0 
\end{equation}
and
\begin{equation} \label{friedmann axion}
 \Bar{\mfa}'' +2 \mathcal{H}\Bar{\mfa}'+\frac{2W,_\chi}{W} \, \Bar{\mfa}'\Bar{\chi}'+\frac{a^2 V_{,\mfa}}{\MPL ^2 W^2}  =0\, ,
\end{equation}
where $\Bar{\rho} = \Bar{\rho}_\ssB + \Bar{\rho}_{\ssR}$, the Hubble expansion rate is $\mathcal{H} = a'/a$ and primes denote differentiation with respect to conformal time. 

The conservation equation for the background energy in radiation implies $\bar\rho_{\ssR} \propto a^{-4}$ as usual, while conservation of energy for the background baryon distribution ensures  
\begin{equation}\label{contequ B}
 \bar{\rho}'_{\ssB}+3\mathcal{H}\Bar{\rho}_{\ssB}=\Bar{\rho}_{\rm \ssB}  \mathbf{g} \, \Bar{\chi}' \,,
\end{equation}
which is consistent with $\bar \rho_\ssB \propto m(\bar\chi)/a^3$ where $m(\bar \chi) = m \, e^{\,\bfg \,\bar\chi}$ captures the universal dilaton dependence of ordinary particle masses implied by the dilaton-matter coupling. In practice it is the nucleon mass\footnote{Because the nucleon mass dominantly consists of gluons one might think it need not scale with the dilaton as do other elementary Standard Model particles. Computing the QCD scale using the renormalization group, however, shows that it is proportional to the UV scale where the value of the coupling is chosen (such as at the $Z$-boson mass). The nucleon mass then scales universally with the dilaton if this UV scale also does so (as it would if it were the $Z$ mass) \cite{Burgess:2021obw}.} that dominates in the sum over nonrelativistic species.

\subsubsection*{Perturbations}

To compute cosmic fluctuations we perturb the field equations about these homogeneous background configurations and keep only terms linear in the fluctuations, working in conformal-Newtonian gauge. We furthermore assume only scalar metric perturbations
\begin{equation}
    {\rm d}s^2 = a^2(\eta)\Bigl[-(1+2\Phi){\rm d}\eta^2 + (1-2\Psi) \delta_{ij} \, {\rm d}x^i {\rm d}x^j\Bigr] \,,
\end{equation}
and perturbed components of the fluid 4-velocity given by 
\begin{equation}\label{4VelocityExpansion}
u^\mu_f \partial_\mu = \frac{1}{a} (1- \Phi) \partial_\eta + \frac{ \partial^i v_f}{a} \partial_i \, ,
\end{equation}
for each fluid (where $\partial_\eta = \partial/\partial \eta$). The perturbed energy densities and scalar fields are  
\begin{equation}
\rho_\ssB \equiv \bar\rho_\ssB (1+\delta_\ssB) \, , \quad \rho_\ssR \equiv \bar\rho_\ssR (1+\delta_\ssR) \, , \quad \chi \equiv  \bar\chi + \delta\chi \qq{and}  \mfa \equiv\bar{\mfa}+\delta\mfa \, ,
\end{equation}
and we assume the absence of anisotropic stress. 

Going to Fourier space, the perturbed $(\eta,\eta)$ component of the Einstein equation becomes
\begin{align}
\label{Perturbed Friedmann}
& k^2\Psi+3\mathcal{H}\Psi'+ \frac{1}{2}\Bigl( \Bar{\chi}'\delta\chi'+W^2\Bar{\mfa}'\delta\mfa'\Bigr) +\frac{1}{2} W\, W,_\chi \Bar{\mfa}'^2\delta\chi \nonumber
\\
&\qquad \qquad +\frac{a^2}{2\MPL ^2}\Bigl( 2\Phi V + V,_{\bar{\chi}}\delta\chi + V,_{\Bar{\mfa}}\delta\mfa\Bigr) =-\frac{a^2}{2\MPL ^2} \left(\delta\rho+2\Phi\bar{\rho}\right) \,,
\end{align}
where $\rho$ without a subscript denotes the total fluid energy density: $\rho =\sum_f \rho_f$. The perturbed $(\eta,i)$ components of the Einstein equations (\ref{einstein eq}) similarly are
\begin{equation} \label{perturbed 0-i}
 k^2(\Psi'+\mathcal{H}\Phi)-\frac{k^2}{2}\Bigl( \Bar{\chi}'\delta\chi+W^2\Bar{\mfa}'\delta\mfa\Bigr)=\frac{a^2\Bar{\rho}}{2\MPL ^2}\Theta \, ,
\end{equation}
where $\Theta$ is a sum over the expansion for each fluid component, $\Theta = \sum_f \Theta_f$, with $\Theta_f$ defined for each component by $\Theta_f = -k^2v_f$ where $v_f$ is defined in \pref{4VelocityExpansion}. 
The $(i,j)$ components become
\begin{align} \label{perturbed i-j}
&\Psi''+\mathcal{H}\left(\Phi'+2\Psi'\right)+\left(2\mathcal{H}'+\mathcal{H}^2\right)\Phi+ \frac{1}{2} \Bigl( \Bar{\chi}'^2 + W^2\Bar{\mfa}'^2 \Bigr) \Phi \nonumber
\\
&\qquad\qquad -\frac{1}{2} \Bigl( W^2\Bar{\mfa}'\delta\mfa' + \Bar{\chi}'\delta\chi' + W,_\chi W\Bar{\mfa}'^2 \delta\chi\Bigr)  
+\frac{a^2}{2\MPL ^2} \Bigl( V,_{\Bar{\chi}} \delta\chi + V,_\mfa\delta \mfa\Bigr) = 0 \,.
\end{align}
The quantities $W = W(\bar\chi)$ and $V = V(\bar \chi, \bar \mfa)$ appearing in these equations (and their derivatives) are evaluated at the background. 

The perturbed scalar field equations are
\begin{align}
 \label{perturbed dilaton}
& \delta\chi'' + 2\mathcal{H}\delta\chi' + \left[k^2-\Bar{\mfa}'^2 \left(W,_\chi ^2 + WW,_{\chi\chi}\right) +\frac{a^2}{\MPL ^2} V,_{\Bar{\chi} \Bar{\chi}}\right] \delta\chi-\Bar{\chi}' \left(\Phi'+3\Psi'\right)
 \nonumber
\\
& \qquad\qquad -2WW,_\chi \Bar{\mfa}' \delta\mfa' +\frac{a^2}{\MPL ^2} \Bigl( 2V,_{\chi}\Phi +V,_{\Bar{\chi} \Bar{\mfa}} \delta\mfa\Bigr)   = -\frac{a^2}{\MPL ^2} \, \bfg  \Bigl( \delta_\ssB +2\Phi \Bigr) \Bar{\rho}_\ssB  \, ,   
 \end{align}
and
\begin{align}
\label{perturbed axion}
&\delta\mfa''+\delta\mfa'\left(2\mathcal{H}+2\Bar{\chi}'\frac{W,_\chi}{W}\right)+\left(k^2+\frac{a^2}{\MPL ^2}\frac{V,_{\mfa\mfa}}{W^2}\right)\delta \mfa + \frac{2a^2}{\MPL ^2}\frac{V,_a\Phi}{W^2}+2\Bar{\mfa}'\frac{W,_\chi}{W}\delta\chi'  
\\
&\qquad + \delta\chi \left[2\Bar{\chi}' \Bar{\mfa}'\left(\frac{W,_{\chi\chi}}{W}-\left(\frac{W,_{\chi}}{W}\right)^2\right)+\frac{a^2}{\MPL ^2}\left(\frac{V,_{\Bar{\mfa}\Bar{\chi}}}{W^2}-\frac{W,_\chi}{W^3}V,_{\Bar{\mfa}}\right)\right]
-\left(\Phi' + 3\Psi'\right) \Bar{\mfa}' = 0 \,, \nonumber
\end{align}
while the perturbed continuity equation for baryons is 
\begin{equation} \label{Continuity equation perturbed B}
  \delta_\ssB'+\Theta_\ssB-3\Psi'  = \bfg \delta\chi' \, , 
\end{equation}
and the baryon Euler equations boil down to
\begin{equation}
\label{Euler equation Baryon}
 \Theta'_\ssB+\Theta_\ssB\mathcal{H}-k^2\Phi = - \bfg \left(\Bar{\chi}'\Theta_\ssB-k^2\delta\chi\right)  + J_{{\rm eq}}\, , 
\end{equation}
where \cite{Ma:1995ey}
\begin{equation}
   J_{{\rm eq}} = \frac{4 \bar \rho_\gamma}{3 \bar \rho_\ssB} \, a \, n_e \, \sigma_\ssT \Bigl( \Theta_\gamma - \Theta_\ssB \Bigr) \,,
\end{equation}
describes the energy exchange with the photon fluid where
$\sigma_\ssT$ is the Thomson cross section (including the $\bar\chi$-dependence of the electron mass). 

\subsection{Axion potential and oscillations}
\label{SecIII:MinimalDarkSector}

In order to see whether the axion can be the Dark Matter we next introduce a scalar potential, $V(\mfa,\chi) = V_{\rm ax}(\mfa) + V_{\rm dil}(\chi)$ (where the details of $V_{\rm dil}(\chi)$ are specified below). We imagine the axion to move near the minimum of $V_{\rm ax}$, which we write as
\begin{equation} \label{Vaxeq}
    V_{\rm ax}(\ax) \simeq \tfrac{1}{2} m_\ax^2 \MPL^2 \left(\ax-\ax_0\right)^2.
\end{equation}
It is the damped oscillations about this minimum that ultimately describe Dark Matter, in the limit where the oscillations are much faster than the Hubble time. 

There are two different lines of justification for the use of a quadratic potential like \pref{Vaxeq}. The simplest recognizes that $V_{\rm ax}(\mfa)$ is usually actually a periodic function of $\mfa$ but is well-approximated by a quadratic form for sufficiently small oscillations. In this type of justification it is necessary to verify {\it ex post facto} whether or not oscillation amplitudes are ever large enough for the quadratic approximation to fail. 

The second type of justification for a quadratic potential comes if the axion is generated as the dual of a Kalb-Ramond 2-form field $B_{\mu\nu}$, such as can be the case for extra-dimensional UV completions. In this case the quadratic scalar potential is a direct consequence of working at the two-derivative level in the derivative expansion in the Kalb-Ramond theory (along the lines lines described in \cite{Burgess:2023ifd} based on earlier work in \cite{Quevedo:1996uu}). 

\subsubsection{Adiabatic evolution}

It is too computationally expensive to resolve such fast oscillations numerically, but for longer-time cosmological evolution it suffices to follow the slower evolution obtained after averaging over a large number of oscillation periods. This leads to a formulation for the axion dynamics in which such oscillations are described by an effective axion fluid.

To this end we use the Madelung formalism \cite{Madelung:1927ksh} and explicitly extract the rapid oscillations by writing
\begin{equation} \label{asplit}
    \ax = \mathbf{a}+\frac{1}{\sqrt{2}}\left[\psi \, e^{-i\int \axm(t) \,\exd t}+\psi^* \, e^{i\int \axm(t) \,\exd t}\right], 
    \quad \hbox{where} \quad
     \axm(t) := \frac{m_\ax}{W(\Bar{\s})} \, ,
\end{equation}
is assumed to be super-Hubble: $\mfm \gg H$. The envelope and central value of the oscillations evolve adiabatically as the universe expands, so the remaining variables $\mathbf{a}$ and $\psi$ appreciably evolve only over the much longer Hubble timescale. 

To compute their evolution we first decompose $\psi$ into its modulus and phase, writing
\begin{equation}
    \psi = \frac{1}{\axm(t)} \, \sqrt{\rho_\ax} \; e^{iS} \,,
\end{equation}
and substitute the result into the axion action (\ref{Action}), everywhere dropping $\partial_t\psi$ and $H\psi$ compared with $m(t) \psi$. This leads to
\begin{equation}
     -\sqrt{-g} \left[ \frac12 M_p^2 \, W^2 \partial_\mu \mfa \, \partial^\nu\mfa + V_{\rm ax}(\mfa) \right] 
     = -\sqrt{-g} \left[\frac12 M_p^2 \, W^2 \partial_\mu \bfa \, \partial^\nu\bfa + V_{\rm ax}(\bfa) \right]  +\cL_{\rm fluid},
\end{equation}
where 
%
% \begin{equation}
%     \mathcal{L}_{\rm fluid} := \sqrt{-g} \, W^2 \left\{ \frac{W_{,\s}}{W} \, \delta\s \rho_\ax - \frac{\rho_\ax S'}{a\axm(\eta)} - \Phi \, \rho_\mfa -\frac{1}{2a^2\axm^2(\eta)} \left[ \frac{(\nabla\rho_\mfa)^2}{4\rho_\mfa}+\rho_\mfa(\nabla S)^2\right] \right\}  \,.
% \end{equation}
%
\begin{equation}
    \mathcal{L}_{\rm fluid} := \sqrt{-g} \, \left\{\Bar{W}_{,\s}\Bar{W} \, \delta\s \,\rho_\ax - \frac{\rho_\ax W^2 S'}{a\axm(\eta)} - \Phi \Bar{W}^2 \, \rho_\mfa -\frac{ \Bar{W}^2}{2a^2\axm^2(\eta)} \left[ \frac{(\nabla\rho_\mfa)^2}{4\rho_\mfa}+\rho_\mfa(\nabla S)^2\right] \right\}  \,.
\end{equation}
Here we ignore second order perturbations in the metric and dilaton field as this Lagrangian is exclusively used to calculate the axion equations of motion. The dilaton equation of motion and Einstein equations can be derived from (\ref{Action}), and the axion terms can be averaged to give the fluid contribution in each of these equations. 

Computing the Euler-Lagrange equations shows that $\rho_\mfa$ satisfies a fluid-like equation 
\begin{equation} \label{fluidEq}
\frac{1}{W^2}\partial_\eta\left(\left[1-\Phi-3\Psi\right]{W^2 {\rho}_\mfa}\right)+ 3 H \rho_\mfa -\frac{\axm'}{\axm}\rho_\mfa+ \frac{1}{a} \rho_\mfa\vec (\nabla\cdot \vec v_\mfa)=0,
\end{equation}
where the fluid velocity is defined by $\vec v_a \equiv { \nabla S}/[{m(\eta) a}]$. The idea is to compute the axion contribution to the time-averaged equations over long timescales by tracking the background and fluctuation parts of these fluid equations. In particular, the background homogeneous solution to \pref{fluidEq} gives 
\begin{equation}\label{Fluid evol}
    \Bar{\rho}_\sax := \Bar{W}^2\Bar{\rho}_\ax = \frac{C\axm(\eta)}{a^3} ,
\end{equation}
where $C$ is a proportionality constant (whose value is chosen to ensure the energy density of the axion fluid has the right size to be the Dark Matter\footnote{In practice we set the constant $C$ by ensuring that the fractional share of Dark Matter today $\Omega_{\textnormal{ax}0}h^2 = 0.120$ agrees with the Planck 2018 best-fit value taken from \cite{Planck:2018vyg}, where $h = 0.674$, and return to the issue of physical production mechanisms responsible for setting this constant in \S\ref{sec:AxionProperties}\;. }).  As expected, we find the standard $a^{-3}$ falloff for the energy of the fast oscillations, supplemented by the time-dependence of $\axm$.

The zero-mode field $\bfa$ similarly evolves according to the axion Klein-Gordon equation
\begin{equation}
    \bax''+2\mathcal{H}\bax' + \frac{2W_{,\s}}{W}\bax'\s' + \frac{a^2}{W^2}V_{,\bax} = 0.
\end{equation}
Because fast oscillations are not part of the definition of $\bfa$ we solve this equation by setting $\bfa$ at the potential's minimum, $\bax = \ax_0$, which implies $\bfa'' =\bfa' = 0$ if the potential's parameters are time-independent (as they are here). It is $\psi$ that parameterises the oscillations around this minimum, and the variable $\rho_\mfa$ that captures its slow evolution.

Using these expressions allows the axion evolution with the fast oscillations averaged out to be described in terms of the effective fluid. 
%
%This contributes to the Friedmann equation \Cliff{This was written $H$ as if we were writing it using cosmic time - I started converting to $\cH$ and conformal time, but am not sure if this is a typo} 
%
%\begin{equation}\label{Friedmann eq}
%    \cH^2 = \rho + \frac{\s'^2}{2}+V_{\rm dil}(\s) + W^2\rho_\mfa \,,
%\end{equation}
%
%showing how the other fluids react to the axion energy density. The background dilaton equation similarly becomes
%
%\begin{equation}\label{Dilaton eq}
%    \s''+2\mathcal{H}\s'+V_{{\rm dil},\s} = W \, W_{,\s} \rho_\mfa-\mathbf{g}\rho_\ssB.
%\end{equation}
%
Averaging the metric fluctuation equations \pref{Perturbed Friedmann} through \pref{perturbed i-j} over many axion oscillations leads to the expressions
%
% \begin{align}
% \label{Perturbed Friedmann2}
% & k^2\Psi+3\mathcal{H}\Psi'+ \frac{1}{2}\Bigl( \Bar{\chi}'\delta\chi'+\frac{W^2}{2}\delta\left<\mfa'^2\right>\Bigr) +\frac{1}{2} W\, W,_\chi \left<\Bar{\mfa}'^2\right>\delta\chi \\
% &\qquad \qquad\qquad \qquad +\frac{a^2}{2\MPL ^2}\Bigl( 2\Phi V + V,_{\bar{\chi}}\delta\chi + \delta\left<V(\mfa)\right>\Bigr) =-\frac{a^2}{2\MPL ^2} \left(\delta\rho+2\Phi\bar{\rho}\right) \, , \nonumber
% \end{align}
%
\begin{align}
\label{Perturbed Friedmann2}
& k^2\Psi+3\mathcal{H}\Psi'+ \frac{1}{2}\Bigl( \Bar{\chi}'\delta\chi'+\frac{1}{2}\delta\left<\Bar{W}^2\mfa'^2\right>\Bigr) +\frac{1}{2} \frac{W,_\chi}{W}\left<\Bar{W}^2\Bar{\mfa}'^2\right>\delta\chi \\
&\qquad \qquad\qquad \qquad+\frac{a^2}{2\MPL ^2}\Bigl( 2\Phi V + V,_{\bar{\chi}}\delta\chi + \delta\left<V(\mfa)\right>\Bigr) =-\frac{a^2}{2\MPL ^2} \left(\delta\rho+2\Phi\bar{\rho}\right) \, ,\nonumber
\end{align}
%
% \begin{equation}
% \label{perturbed 0-i2}
%  k^2(\Psi'+\mathcal{H}\Phi)-\frac{k^2}{2} \Bar{\chi}'\delta\chi+\frac{W^2}{2}\left<k^i\partial_i \mfa\partial^0\mfa\right>=\frac{a^2\Bar{\rho}}{2\MPL ^2}\Theta \, ,
% \end{equation}
%
\begin{equation}
\label{perturbed 0-i2}
 k^2(\Psi'+\mathcal{H}\Phi)-\frac{k^2}{2} \Bar{\chi}'\delta\chi+\frac{1}{2}\left<W^2 k^i\partial_i \mfa\partial^0\mfa\right>=\frac{a^2\Bar{\rho}}{2\MPL ^2}\Theta \, ,
\end{equation}
and
%
% \begin{align}
% \label{perturbed i-j2}
% &\Psi''+\mathcal{H}\left(\Phi'+2\Psi'\right)+\left(2\mathcal{H}'+\mathcal{H}^2\right)\Phi+ \Bigl( \Bar{\chi}'^2 + W^2\left<\Bar{\mfa}'^2\right> \Bigr) \Phi \\
% &\qquad\qquad -\frac{1}{2} \Bigl( \frac{W^2}{2}\delta\left<\mfa'^2\right> + \Bar{\chi}'\delta\chi' + W,_\chi W\left<\Bar{\mfa}'^2\right> \delta\chi\Bigr)  
% +\frac{a^2}{2\MPL ^2} \Bigl( V,_{\Bar{\chi}} \delta\chi + \delta\left<V(\mfa)\right>\Bigr) = 0 \,.\nn
% \end{align}
%
\begin{align}
\label{perturbed i-j2}
&\Psi''+\mathcal{H}\left(\Phi'+2\Psi'\right)+\left(2\mathcal{H}'+\mathcal{H}^2\right)\Phi+ \Bigl( \Bar{\chi}'^2 + \left<\Bar{W}^2\Bar{\mfa}'^2\right> \Bigr) \Phi 
\\
&\qquad\qquad -\frac{1}{2} \Bigl( \frac{1}{2}\delta\left<\Bar{W}^2\mfa'^2\right> + \Bar{\chi}'\delta\chi' + \frac{W,_\chi}{W}\left<\Bar{W}^2\Bar{\mfa}'^2\right> \delta\chi\Bigr)  
+\frac{a^2}{2\MPL ^2} \Bigl( V,_{\Bar{\chi}} \delta\chi + \delta\left<V(\mfa)\right>\Bigr) = 0 \,.\nn
\end{align}
Averaging the perturbed dilaton equation \pref{perturbed dilaton} similarly gives
%
% \begin{align}
%  \label{perturbed dilaton2}
% & \delta\chi'' + 2\mathcal{H}\delta\chi' + \left[k^2-\left<\Bar{\mfa}'^2 \right>\left(W,_\chi ^2 + WW,_{\chi\chi}\right) +\frac{a^2}{\MPL ^2} V,_{\Bar{\chi} \Bar{\chi}}\right] \delta\chi-\Bar{\chi}' \left(\Phi'+3\Psi'\right)
% \\
% & \qquad\qquad \qquad\qquad \qquad\qquad\qquad  -WW,_\chi \delta\left<{\mfa}'^2\right> +2\frac{a^2}{\MPL ^2} V,_{\chi}\Phi 
%   = -\frac{a^2}{\MPL ^2} \, \bfg \Bigl( \delta_\ssB +2\Phi \Bigr) \Bar{\rho}_\ssB  \, . \nn
% \end{align}
%
\begin{align}
& \delta\chi'' + 2\mathcal{H}\delta\chi' + \left[k^2-\left<\Bar{W}^2\Bar{\mfa}'^2\right>\left(\frac{W,_\chi ^2}{W^2} + \frac{W,_{\chi\chi}}{W}\right) +\frac{a^2}{\MPL ^2} V,_{\Bar{\chi} \Bar{\chi}}\right] \delta\chi-\Bar{\chi}' \left(\Phi'+3\Psi'\right)
\\
& \qquad\qquad \qquad\qquad \qquad\qquad\qquad  -\frac{W,_\chi}{W} \delta\left<\Bar{W}^2\mfa'^2\right> +2\frac{a^2}{\MPL ^2} V,_{\chi}\Phi 
  = -\frac{a^2}{\MPL ^2} \, \bfg \Bigl( \delta_\ssB +2\Phi \Bigr) \Bar{\rho}_\ssB  \, . \nn
\end{align}
The averaged terms appearing in these equations can be evaluated in terms of fluid contributions and $\bfa$ using  
%
% \begin{align}
% &\delta\left<\mfa'^2\right>=a^2\delta\rho_\mfa-2a\Bar{\rho}_\mfa\frac{ S'}{m(\eta)} \,, \qquad    \left<\Bar{\mfa}'^2\right>=\mathbf{a}'^2+a^2\Bar{\rho}_a,
%     \\
%     &\; \delta\left<V(\mfa)\right>= \frac{m_\mfa^2}{2}\frac{\delta\rho_\mfa}{\axm^2(\eta)} \qq{and} \left<k^i\partial_i \mfa\partial^0\mfa\right> = -\frac{\Bar{\rho}_\mfa}{\MPL^2}\Theta_\mfa\,.
% \end{align}
%
\begin{align}
&\delta\left<\Bar{W}^2\mfa'^2\right>=a^2\Bar{\rho}_{\sax}\delta_\mfa-2a\Bar{\rho}_\sax\frac{ S'}{m(\eta)} \,, \qquad    \left<\Bar{W}^2\Bar{\mfa}'^2\right>=\Bar{W}^2\mathbf{a}'^2+a^2\Bar{\rho}_{\sax},
    \\
    &\; \delta\left<V(\mfa)\right>= \frac{\Bar{\rho}_{\sax}}{2}\delta_\mfa \qq{and} \left<W^2 k^i\partial_i \mfa\partial^0\mfa\right> = -\frac{\Bar{\rho}_{\sax}}{\MPL^2}\Theta_\mfa\,.
\end{align}
%
% \textcolor{red}{where $\delta\rho_\sax \equiv -\delta T_0^0 = \Bar{W}^2\delta\rho_\mfa - \Phi\Bar{W}^2\Bar{\rho}_\mfa$}.
The evolution of the axion fluid fluctuations descend from the axion field equation and are given by\footnote{We remind the reader that $\delta_\mfa$ is the density contrast of $\rho_\mfa$ and not of the physical density as defined by the time-time component of the stress--energy momentum tensor, see also eq. (\ref{Fluid evol}). It can be shown that in the Newtonian gauge the relationship between the density contrast $\delta_\mfa$ and the physical density contrast $\delta_f$ is given by $\delta_f = \delta_\mfa -\Phi - \Phi_\phi$, where $\Phi_\phi \equiv -\frac{W,_\chi}{W}\delta\chi$. We refer to \cite{screenedcosmo}, where details of the calculations can be found.}
\begin{equation}
    \delta_\mfa'-2\Phi_\phi' + \Theta_\mfa = 3\Psi'+\Phi'
\end{equation}
%
% \begin{equation}
%     \textcolor{red}{\delta_\sax'-2\Phi_\phi' + \Theta_\mfa = 3\Psi'}
% \end{equation}
%
% \textcolor{red}{Where $\delta_\sax = \delta_\mfa - \Phi$},
and 
\begin{equation}\label{phase equation}
    \frac{S'}{a\axm(\eta)} = - \Phi_\phi - \Phi_\ssQ -\Phi \,,
\end{equation}
where we define (to linear order)
%
%\begin{equation}
%    \Phi_\ssQ :=  -\frac{1}{W^2(\phi)}\vec \nabla \left(W^2(\phi) \frac{\vec \nabla \rho_\mfa}{4a^2\axm^2(t)\rho_\mfa }\right)-\frac{(\nabla\rho_\mfa)^2}{8a^2m^2(t)\rho_\mfa^2},
%\end{equation}
%
\begin{equation}
    \Phi_\ssQ := \frac{k^2\delta_\mfa}{4a^2\axm^2(\eta)}
\quad \hbox{and}\quad
\Phi_\phi := -\frac{W_{,\chi}}{W}\delta\chi.
\end{equation}
Taking $\nabla^2$ of (\ref{phase equation}) and using the definition $\vec v_a \equiv { \nabla S}/[{m(\eta) a}]$ of the axion fluid velocity allows us to write the axion Euler equation as 
\begin{equation}
    \Theta_\mfa' + \left(\mathcal{H} + \frac{\axm'}{\axm}\right)\Theta_\mfa = k^2\left(\Phi+\Phi_\phi + \Phi_Q\right) \,,
\end{equation}
Finally, the baryons evolve as they would in a single field dilaton model,
\begin{equation} \label{baryonevo}
     \delta_\ssB'+\Theta_\ssB-3\Psi'  = \mathbf{g}  \delta\chi', \qq{and} \Theta'_\ssB+\Theta_\ssB\mathcal{H}-k^2\Phi=- \mathbf{g}  \left(\Bar{\chi}'\Theta_\ssB-k^2\delta\chi\right).
\end{equation}
These expressions -- eqs.~\pref{Perturbed Friedmann2} through \pref{baryonevo} -- are the equations to be evolved numerically in later sections, although in the code we implement these equations in the synchronous gauge. But before doing so we must first specify our choices for $W(\chi)$ and $V_{\rm dil}(\chi)$.

\subsection{Dilaton properties}
\label{sec:DilatonProperties}

As mentioned in the introduction, approximate scaling symmetries suggest our choices for $W$ and $V_{\rm dil}$ should be exponentials (or approximately so), as given in \pref{ExponentialChoices} and \pref{JFtoEF} (and repeated here for convenience):
\begin{equation} \label{ExponentialChoicesAgain}
    W(\chi) = W_0 \, e^{-\zeta \chi}  \,, \quad
    V_{\rm dil}(\chi) = V_0 \, e^{- \lambda \chi}
    \quad \hbox{and} \quad
    \tilde g_{\mu\nu} := e^{2 \bfg \chi} \, g_{\mu\nu} \,.
\end{equation}
Because $W_0$ can be set to unity by shifting $\chi$ in principle the free parameters of the dilaton sector of the model then are $\zeta$, $\lambda$, $\bfg$ and the coefficient $V_0$.

\subsubsection{Coupling relations}

In practice microscopic models that produce such scalars tend also to predict relationships amongst these parameters; we list a few that help motivate the models we explore. 

\medskip\noindent
\emph{Scale invariance:} 

\medskip\noindent
Notice that an exponential form for $W$ automatically assures the kinetic terms
\begin{equation}
    \cL_{\rm kin} = - \tfrac12 \, \MPL^2 \sqrt{-g} \; g^{\mu\nu} \Bigl( R_{\mu\nu} + \partial_\mu \chi \, \partial_\nu \chi + e^{-2\zeta \chi} \, \partial_\mu \mfa \, \partial_\nu \mfa \Bigr)
\end{equation}
scale by a common factor $\cL_{\rm kin} \to e^{\beta c} \cL_{\rm kin}$ when $\chi \to \chi + c$, $\mfa \to e^{\zeta c} \mfa$ and $g_{\mu\nu} \to e^{\beta c} g_{\mu\nu}$ for arbitrary constants $\beta$, $\zeta$ and $c$. But if we demand a potential term like $\cL_{\rm pot} = - \sqrt{-g} \; V_{\rm dil}$ to scale the same way we must ask $\lambda = \beta$. Similarly, if the scalar potential arises as a constant Jordan-frame potential of the form $\cL_{\rm cc} = - \mu^4 \sqrt{- \tilde g}$ for constant $\mu$ then $\cL_{\rm cc} = - \mu^4 e^{-4\bfg \chi} \sqrt{-g}$ and so $\lambda = 4 \bfg$. This then scales the same way as the kinetic terms only if $\bfg = \frac14 \beta$.

\medskip\noindent
\emph{Supersymmetry:} 

\medskip\noindent
Supergravity provides another broad class of well-motivated theories that predict relationships between the parameters in \pref{ExponentialChoicesAgain}. In these models the dilaton and axion arise as real and imaginary parts of a complex scalar $T = \tau + i \mfa$. For these the function $W(\chi)$ and the relationship between $\tilde g_{\mu\nu}$ and $g_{\mu\nu}$ are related to each other because both can often be derived from the same K\"ahler potential. For instance if $K = - 3 \alpha \log(T+T^*)$ then one finds the Einstein-frame kinetic term is proportional to
\begin{equation}
     K_{TT^*} \partial_\mu T^* \, \partial^\mu T =  \frac{3\alpha}{4\tau^2} \Bigl[ \partial_\mu \tau \, \partial^\mu \tau + \partial_\mu \mfa \, \partial^\mu \mfa \Bigr]
\end{equation}
and so $\tau \propto e^{\zeta \chi}$ achieves our assumed normalization for $\chi$ if $\zeta = \sqrt{2/(3\alpha)}$. But Einstein frame is also often achieved in the supergravity action after Weyl rescaling the metric by $\tilde g_{\mu\nu} = e^{K/3} g_{\mu\nu}$ (in Planck units)
and so if the initial metric were the one to which matter couples then $\bfg = - \frac12 \alpha \zeta = - \sqrt{\alpha/6}$. Under broad assumptions supersymmetric models also predict a scalar potential for $\tau$ that is proportional to $(\alpha-1)e^K$ corresponding to a potential of the form \pref{ExponentialChoicesAgain} with $\lambda = 3\alpha$ and $V_0 \propto (\alpha-1)$. The special case where $\alpha = 1$ is a specific example of the `no-scale' form \cite{Cremmer:1983bf, Barbieri:1982ac, Chang:1983hk, Ellis:1983sf, Barbieri:1985wq}).

\medskip\noindent
\emph{Yoga models:} 

\medskip\noindent
Yoga models \cite{Burgess:2021obw} are a specific class of theories that combine both types of predictions. They are a supersymmetric example of no-scale form, and so $\alpha = 1$ which implies $\zeta = \sqrt{\frac23}$ and $\bfg = - \frac12 \zeta = -\sqrt{\frac16}$. 

The model is designed so that the leading scale-invariant part of the potential vanishes and so the dominant term comes from scale-breaking effects. In particular it turns out that instead of \pref{ExponentialChoicesAgain} the scalar potential has the form
\begin{equation} \label{YogaPot}
    V_{\rm dil}(\chi) = U(\chi) \, e^{-4 \zeta \chi} \qquad \hbox{(Yoga example)}
\end{equation}
and so $\lambda = 4\zeta$. Scale breaking also permits the prefactor $U$ to depend on $\chi$ (which allows $V_{\rm dil}$ to have a minimum for finite $\chi$). In practice, when required to use a specific form for $U$ we choose 
\begin{equation} \label{Uquad}
     U(\chi) = V_0\left[1-u_1 \, \chi+\frac{u_2}{2} \, \chi^2\right] \,,
\end{equation}
with $V_0 \sim \MPL^4$ and the coefficients $u_i$ chosen to allow a nontrivial minimum for $V_{\rm dil}$. The constants $u_i$ are chosen to be order 50 in size so that the potential is minimized for $\chi_{\rm min} \sim \cO(60)$. 

These choices are driven by phenomenology since they ensure $\tau := e^{\zeta\chi} \sim \MPL^2/\MEW^2 \sim 10^{28}$ and this in turn gives the successful order of magnitude $V_{\rm min} \sim (\MEW^2/\MPL)^4$. Assuming $\MPL$ is also the only scale in the matter lagrangian $\cL_m$ then implies the Higgs vev (in Einstein frame) is order $v \sim M_p/\sqrt{\tau} \sim \MEW$ and so the same is also true for most ordinary particle masses (see \cite{Albrecht:2001xt, Burrage:2018dvt} for precursors of this type of framework). The exception is neutrinos, whose masses are quadratic in the Higgs vev and so are $m_\nu \sim M_p / \tau \sim \MEW^2/\MPL$ in size. So the one value of $\chi$ at the potential's minimum simultaneously sets the size of the electroweak scale, neutrino masses and the Dark Energy density. The models also have supersymmetric large extra dimensions \cite{Aghababaie:2003wz, Brax:2022vlf} as a plausible UV completion.

The big question is whether these choices can be made in a technically natural way, and Yoga models explore a relaxation mechanism aimed at ensuring this is true -- see \cite{Burgess:2021obw} for details. The large size of $\bfg$ provides the model's biggest challenge because this is at face value ruled out by solar system tests (see below), making the search for screening mechanisms the main current focus within this framework \cite{Brax:2023qyp}.

\subsubsection{Non-cosmological constraints}

In this paper we take a phenomenological point of view and assume the constants $\lambda$ and $\zeta$ are related by $\lambda = 4 \zeta$, so
\begin{equation}\label{pure exp pot}
     W(\s) = e^{-\zeta\s} \qq{and} V(\chi) = V_0 \, e^{-4\zeta \s}. 
\end{equation}
since the Yoga experience suggests that this leads to a successful description of the hierarchy of scales (and in particular gives acceptably large particle masses and an acceptably small potential energy if $\chi \sim 60$). We part company with Yoga models, however, by demanding that the dilaton-matter coupling be consistent with tests of gravity within the solar system without the need for screening. This requires us to choose $\abs{\mathbf{g}}\lesssim 2\times10^{-3}$ \cite{Will:2014kxa, Bertotti:2003rm}, independent of the value chosen for $\zeta$ (which we regard to be a free parameter).\footnote{The viability of theories (like the Yoga model) with larger values of $\bfg$ requires the existence of a screening mechanism -- such as the one proposed in \cite{Brax:2023qyp} -- that allows the value of $\bfg$ in cosmology to be larger than the one probed in solar-system tests of gravity. For simplicity we do not pursue these options further here.} 

This size for $\bfg$ also ensures mass variations on Earth to be small enough not to have been ruled out so far (but large enough to be worth searching for). For instance a dilaton profile $\chi \sim \bfg GM_\oplus/r$ near the Earth's surface gives a mass difference between two particles situated at altitudes that differ by $h \ll R_\oplus$ of order 
\begin{equation}
    \frac{\Delta m}{m} \sim \frac{\bfg^2 GM_\oplus h}{R_\oplus^2} \sim 10^{-19} \left( \frac{\bfg^2}{10^{-6}} \right) \left( \frac{h}{1 \, \hbox{km}} \right)
\end{equation}
and so lies below the current limits set by atomic clocks. Furthermore the universal Brans-Dicke style coupling to matter through a Jordan-frame metric -- as in \pref{JFtoEF} -- automatically suppresses dilaton-mediated contributions to tests of the equivalence principle \cite{Will:2014kxa}. The most constraining tests like the Cassini bound \cite{Bertotti:2003rm} instead probe whether particles (usually photons) move along geodesics of the Jordan-frame metric $\tilde g_{\mu\nu}$ of \pref{JFtoEF} rather than the Einstein-frame metric $g_{\mu\nu}$.  

We also consider the two cases where $V_{\rm dil}$ involves a nontrivial function $U(\chi)$, as in \pref{Uquad}, or the case where $U = V_0$ is a constant. $U$ can be a constant and still provide a viable description of the Dark Energy equation of state if $\zeta$ is sufficiently small, because slow-roll evolution down a potential proportional to $e^{-\lambda \chi}$ produces an equation of state parameter $w+1 \simeq \frac13 \lambda^2/(1+\frac16 \lambda^2)$. This is consistent with the observational bound $w+1 \lsim 0.1$ on the Dark Energy equation of state provided $\lambda = 4\zeta \lsim 0.5$. But constant $U$ is not viable for $\zeta$ larger than this so in this case a minimum for $U$ is required to allow the potential to dominate the energy density at late times.  

\subsection{Axion properties and production mechanisms}
\label{sec:AxionProperties}

The potential \pref{Vaxeq} introduces the single new parameter $m_a$ into the theory, and this together with the kinetic-term decay constant $f = \MPL W(\bar\chi)$ largely govern the axion's properties. 

These properties turn out to play almost no role in the success or failure of the cosmologies we explore in \S\ref{SecIV:Numerics} below, which only rely on three assumptions: ($i$) that the energy density in oscillations agrees with the Dark Matter density; ($ii$) that oscillation amplitudes are small enough that \pref{Vaxeq} is a good enough approximation; ($iii$) and that oscillations start sometime in the pre-BBN epoch (and so the axion mass satisfies $\mfm(t) \gsim H_{{\rm BBN}} \sim 10^{-17}$ eV).

The precise value of $f$ doesn't play much of a role here because the axion does not couple directly to matter. This choice could have been made differently, such as in \cite{Smith:2024ayu} where axion-matter couplings play an important role (though for which the axion potential was negligible). It is nonetheless worth exploring where $f$ is important, both on general grounds and also to see whether the extremely low values implied by the specific Yoga choices (for which $f \sim \MEW^2/\MPL$) can be acceptable.\footnote{In these models the decay constants relevant to axion couplings to matter can be much larger than $f$ (some examples of this arising from plausible UV completions of this model are explored in  \cite{Brax:2022vlf}).}

In the present instance the values of $f$ and $\mfm$ are relevant to the theory of initial conditions: what sets the initial axion energy density \cite{Preskill:1982cy,Abbott:1982af,Dine:1982ah}? The `misalignment' mechanism is a commonly used way to understand this in axion models, and within this approach producing the observed Dark Matter abundance imposes a relation between $\mfm$ and $f$. For instance when $f$ is smaller than any earlier inflationary scales this relation states (for a review, see \cite{Marsh:2015xka})
\begin{equation}
    \mfm \simeq \frac{6 \Omega_m^2 H_0}{(9 \Omega_r)^{3/2}} \left( \frac{\MPL}{f} \right)^4 \simeq 1 \, \hbox{eV} \left( \frac{10^{11} \, \hbox{GeV}}{f} \right)^4 \,,
\end{equation}
where $H_0$ is the present-day Hubble scale and $\Omega_m$ and $\Omega_r$ are the current fraction of energy in Dark Matter and radiation. This puts $\mfm$ well above $H_{\rm BBN}$ for $f \lsim 10^{15}$ GeV, but also returns super-Planckian masses for $f \lsim 10^4$ GeV.

Such a mechanism clearly cannot apply for the extremely low values predicted by \cite{Burgess:2021obw}, but it is also true that UV physics in such models (such as extra dimensions) also intrudes at the scale $f$, which also requires rethinking naive production mechanisms like vacuum misalignment. It is also true that novel production mechanisms are likely possible, such as if the axion field is generated by an initially rolling $\chi$ field due to the derivative interactions encoded in $W(\chi)$. This is particularly intriguing within the context of inflationary scenarios for this class of models, for which $\chi$ plausibly evolves from much smaller values like $\tau = e^{\zeta\chi} \sim 10^4$ during inflation out to the much larger values required today \cite{Burgess:2022nbx}. 

We here leave the question of axion production as an open question, at least for models for which $f$ is not in the range discussed above.

\section{Cosmological evolution}
\label{SecIV:Numerics}

Once expressed in terms of the axion fluid the background Friedmann and dilaton equations -- eqs.~\pref{Perturbed Friedmann2} and \pref{perturbed dilaton} -- become
\begin{equation} \label{friedman fluid}
    \mathcal{H}^2 = \frac{1}{3\MPL ^2}\left[\frac{\MPL ^2}{2}\Bar{\chi}'^2+a^2V+a^2\Bar{\rho} + a^2\Bar{\rho}_\sax\right] \, ,
\end{equation}
and
\begin{equation}  \label{dilaton fluid}
  \Bar{\chi}''+2\mathcal{H}\Bar{\chi}'+\frac{a^2}{\MPL ^2}V,_{\chi}=\frac{a^2}{\MPL ^2}\Bigl(-\bfg\Bar{\rho}_\ssB +\frac{W,_\chi}{W}\Bar{\rho}_\sax \Bigr) \,.
\end{equation}
These use the definition $\rho_\sax = W^2\rho_\mfa$ given in eq.~(\ref{Fluid evol}) for the axion fluid energy density, which is proportional to $\axm/a^3$. It is the observation that this evolves like nonrelativistic matter when the mass $\axm(t) = m_\mfa/W(\Bar{\chi})$ is roughly constant that suggests exploring whether the axion could itself be Dark Matter.  

This section examines more closely whether identifying the axion as Dark Matter survives closer scrutiny in a unified axio-dilaton dark sector cosmology. To this end we explore the cosmological implications of the axio-dilaton kinetic coupling for Dark Matter/Dark Energy interactions. We can see in particular from (\ref{dilaton fluid}) that the dilaton-baryon and dilaton-axion couplings can compete with one another if the axion fluid density is similar to that of baryons (as it must be if the axion is to be the CDM). 

\subsection{CMB and power spectrum}
\label{sec:CMBPowerSpectrum}

The parameters $V_0$ determining the dark energy scale and the combination $C\axm(t=t_0)$ determining the axion fluid density today are both fixed by ensuring the present-day energy densities be consistent with $H_0 = 100h\; \rm km/s\;M_{pc}^{-1}$ with $h = 0.6756$, $\Omega_{\chi,0}h^2 = 0.31$, $\Omega _{\mfa,0} h^2 = 0.12$ and $\Omega_{\B,0} h^2 = 0.022$.

Once these are fixed our remaining free parameter is $\zeta$, which we scan through a series of representative values. In principle the initial position and velocity of the dilaton field, $\chi_{ini}$ and $\chi_{ini}'$ are also choices, but in practice the initial position -- at the BBN epoch of nucleosynthesis, say -- cannot be chosen very far from its present-day value due to limits on how different particle masses can be back then without ruining the success of BBN predictions. Late-time cosmology is also largely insensitive to the initial dilaton velocity because its subsequent evolution tends to be drawn into a scaling tracker solution \cite{Ferreira:1997hj, Albrecht:2001xt}. In particular, interesting $\chi$ evolution occurs even if $\chi'_{ini} = 0$ at BBN with $\chi_{ini}$ at its present-day value.

\begin{figure}[hbtp!]
    \begin{subfigure}[b]{0.99\textwidth}
        \centering
    \includegraphics[width = \linewidth]{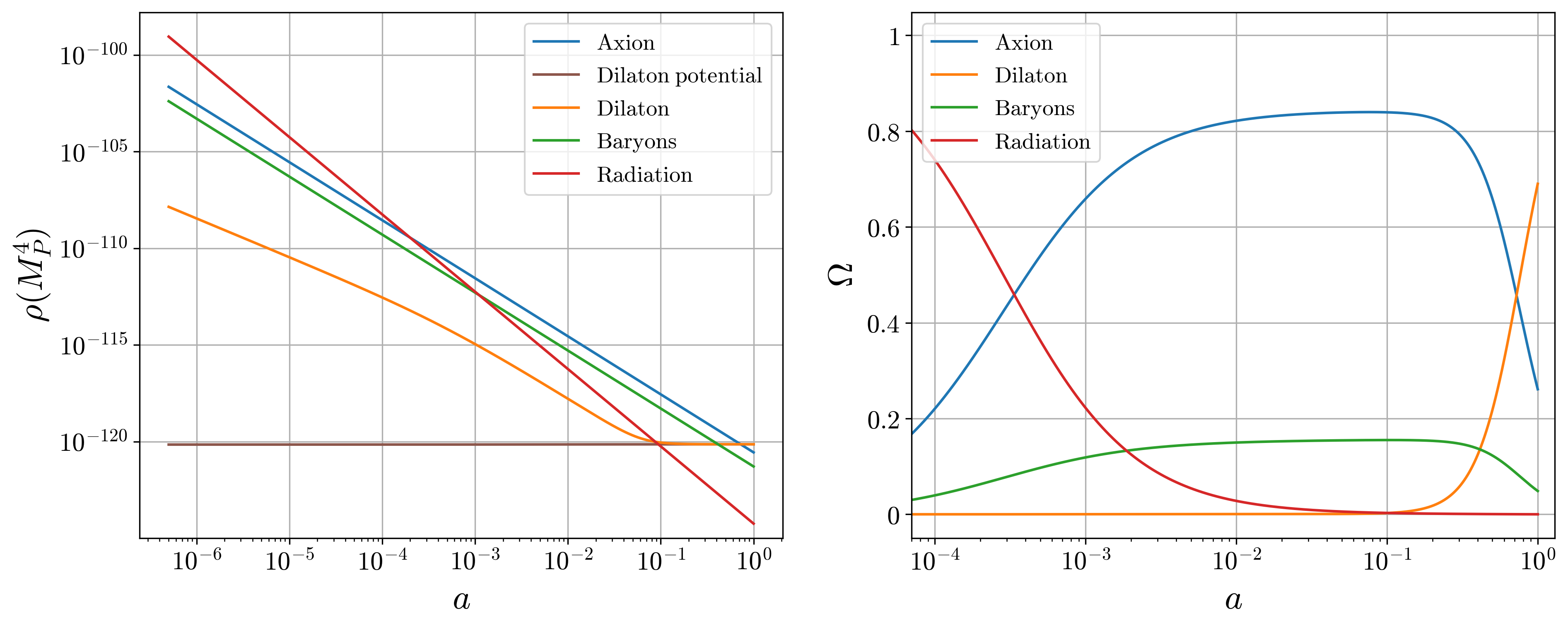}
    \end{subfigure}
    \hfill
    \begin{subfigure}[b]{0.99\textwidth}
        \centering
    \includegraphics[width = \linewidth]{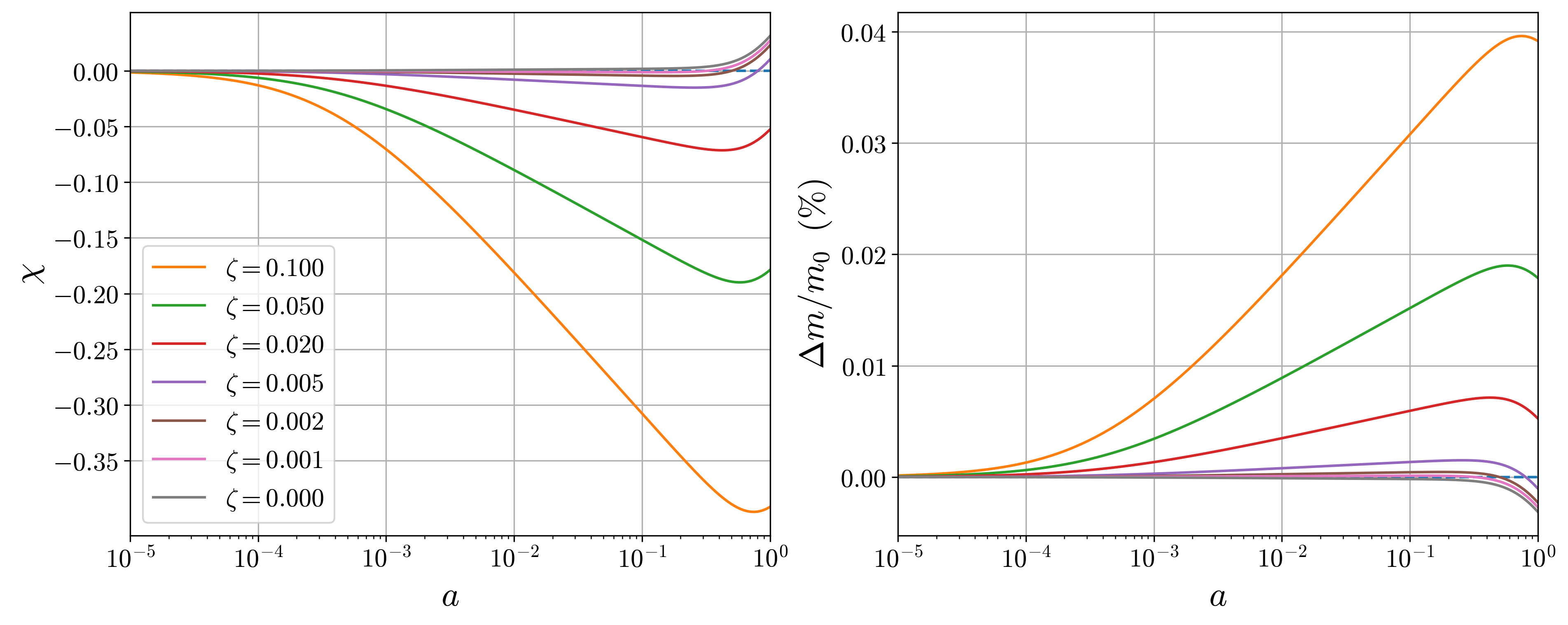}
    \end{subfigure}
    \hfill
    \begin{subfigure}[b]{0.99\textwidth}
        \centering
    \includegraphics[width = \linewidth]{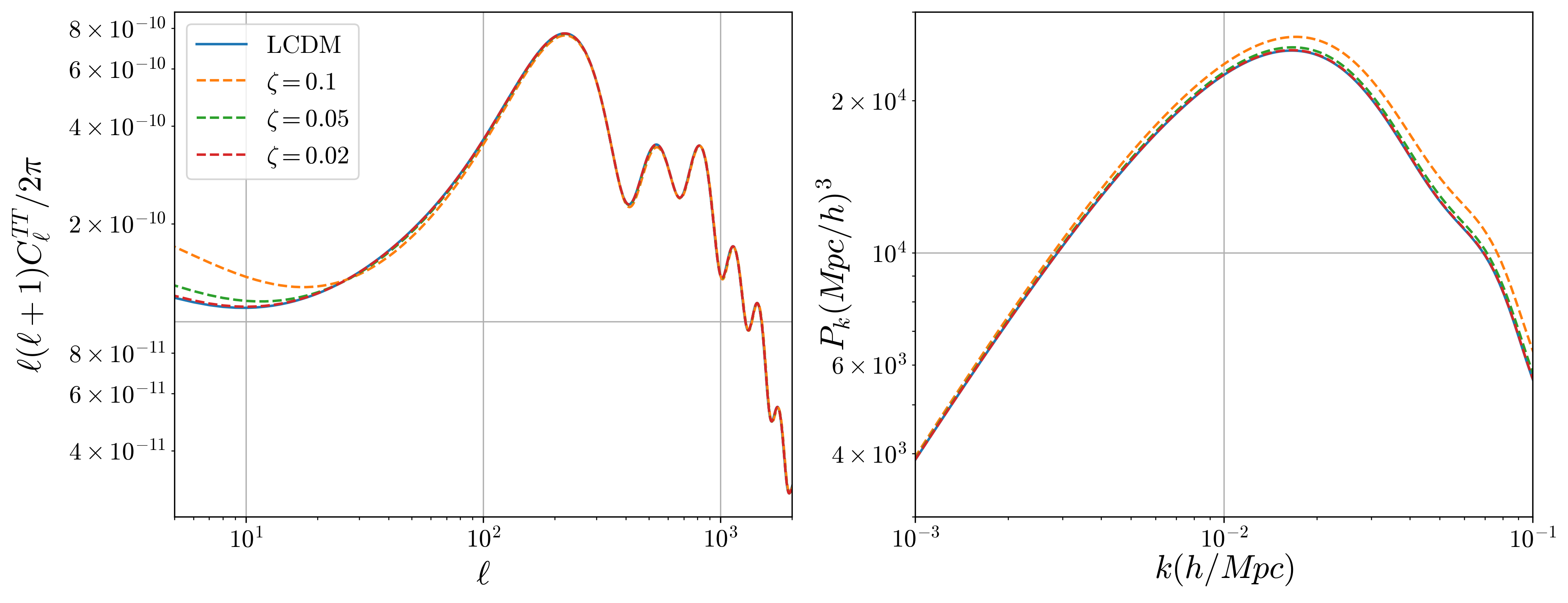}
    \end{subfigure}
    \hfill
    \caption{Background and perturbative level plots of the axio-dilaton cosmology with a pure exponential dilaton potential using CLASS \cite{Diego_Blas_2011}. Top row shows the evolution of the relevant background energy densities when $\zeta = 0.1$. Middle row shows the evolution of the dilaton field and associated baryon masses for a range of $\zeta$. Bottom row shows the corresponding angular and matter power spectra with the $\Lambda\rm CDM$ best-fit in solid blue. In all cases $\mathbf{g}=-10^{-3}$. }
    \label{fig:No well}
\end{figure}

Figure \ref{fig:No well} shows what happens when we set $u_1 = u_2 = 0$ in (\ref{Uquad}) and $\mathbf{g} = -10^{-3}$ and vary $\zeta$, with the dilaton's position at BBN chosen to agree with its present-day value. The plots in the first row of this figure show the energy densities of the background fluids and these only depend on $\zeta$ when it is much larger than the range we consider here. The curves in the top left plot are mostly straight lines on a log-log plot because the energy densities all fall as a power of $a$. As mentioned above, the dilaton does so in particular because the scale-invariance of the exponential interactions drives the fields into scaling `tracker' solutions whose form depends on how the total energy density scales with universal expansion. This tracker nature is seen in the evolution of dilaton energy, whose slope changes with the slight bend at radiation-matter equality.

The second row of the figure shows how $\chi$ evolves and the associated size of the fractional change in ordinary particle masses that this produces. This evolution has several noteworthy features.  The first observation is that $\chi$ does not actually evolve very far before radiation-matter equality despite being described there by a scale invariant tracker solution. This is because of the ruthlessness of Hubble friction in the early universe where radiation dominates the dilaton energy density \cite{Albrecht:2001xt}. 

The second observation is that $\chi$ generically undergoes an excursion around radiation-matter equality even if started off at rest at earlier times. This excursion -- observed also in \cite{Burgess:2021obw} and \cite{Smith:2024ayu} -- is a robust transient feature of these cosmologies caused by the transition between the different tracker solutions appropriate for radiation and matter dominated universes. 

The third noteworthy observation is that when $u_1 = u_2 = 0$ the results are largely insensitive to the value chosen for the constant $W_0$, and this also makes them insensitive to the precise value assumed for the axion decay constant. This independence is both found from explicit numerical evolution and can be seen from evolution equations \pref{Perturbed Friedmann2} through \pref{baryonevo}, which mostly depend on the logarithmic derivative of $W$ once the axion contributions are expressed in terms of $\rho_{\rm ax}$. The size of $W_0$ only starts to play a visible role once the `quantum pressure' $\Phi_\ssQ$ becomes important (and so only for extremely light axions).

The second row of plots also show that larger values of $\zeta$ cause larger mass excursions, with $\chi$ driven preferentially towards smaller values (so particle masses are driven to larger values). This can be understood because the tracker solutions depend on the argument in the exponential effective potential seen by the dilaton. This can depend on $\zeta$ because the $W$-dependent last term on the right-hand side of \pref{dilaton fluid} converts axion oscillations into a contribution to the dilaton potential. For large enough $\zeta$ this can be the most important force seen by $\chi$, particularly during the axion-dominated part of the cosmological evolution. Unlike the actual scalar potential $V(\chi)$ and the dilaton coupling to matter the axion-generated potential pushes the dilaton to \emph{smaller} values of $\chi$, and so causes particle masses to increase.  The size of the resulting mass change is controlled by the size of $\bfg$.

\begin{figure}[hbtp!]
    \begin{subfigure}[b]{0.99\textwidth}
        \centering
    \includegraphics[width = \linewidth]{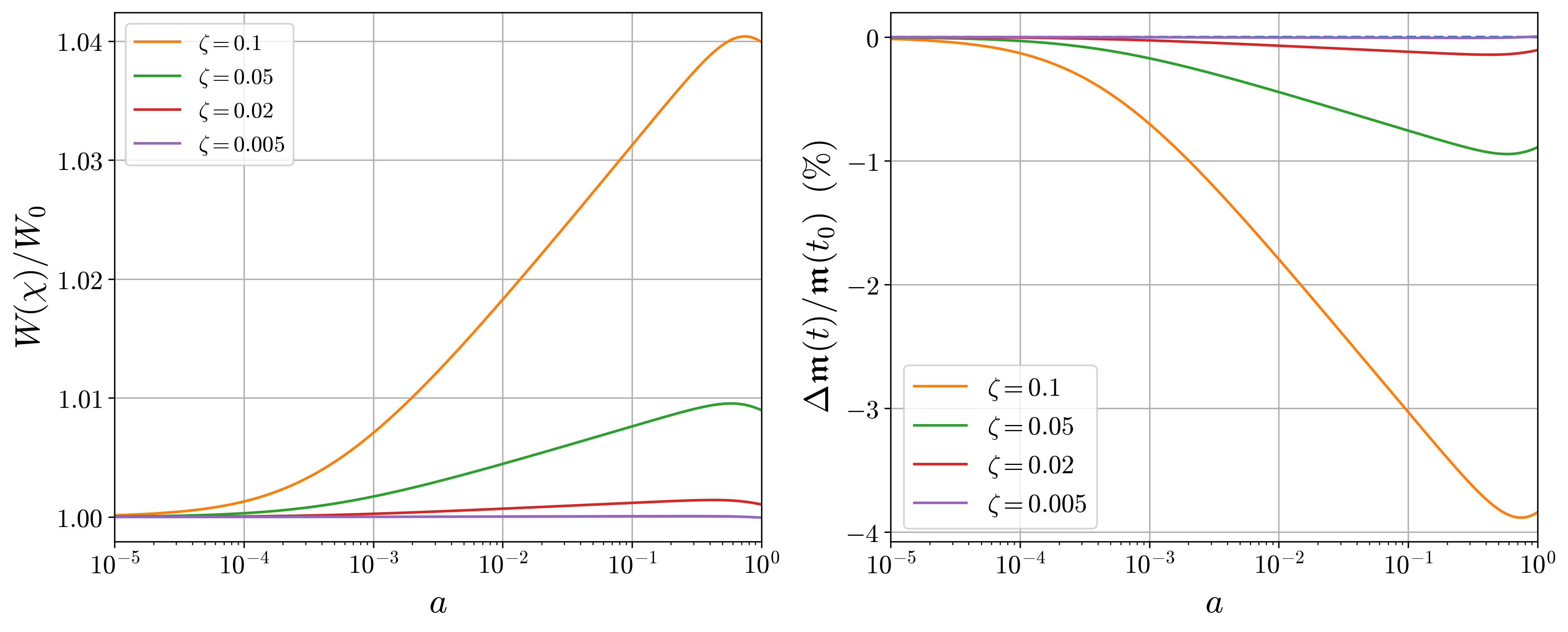}
    \end{subfigure}
    \hfill
    \caption{Fractional changes in the axion decay constant $W(\chi)$ and effective mass $\mfm(t)$ with scale factor in the case of an exponential dilaton potential.}
    \label{fig:No well W and axion mass}
\end{figure}

The third row of the figure shows the implications of the model for the CMB and the matter power spectrum. These plots show how increased dilaton evolution also has implications here because it affects matter clustering, for which the dominant deviations from $\Lambda$CDM arise due to the axion -- it is the dark matter after all -- which is allowed to couple to the dilaton much more strongly than do ordinary particles. This stronger coupling also makes the axion mass $\axm$ more sensitive to excursions in $\bar\chi$ -- and so also $\zeta$ -- through \pref{asplit}, as shown in fig. \ref{fig:No well W and axion mass}.  

Variations of the axion mass due to dilaton evolution also causes the axion (Dark Matter) density to deviate from $1/a^3$ evolution -- {\it c.f.}~eq.~\pref{Fluid evol} --  resulting in greater variations in the Newtonian potentials between recombination and today. This induces larger and larger integrated-Sachs-Wolfe (ISW) effects on CMB photons. For large $\zeta$ these are responsible for the deviations from the $\Lambda$CDM best-fit for small multipoles ($5 \lsim \ell\lsim 15$) seen in the angular power spectrum of the bottom left panel of figure \ref{fig:No well}.

Further reducing $\zeta$ eventually leads to the dilaton-baryon coupling term in (\ref{dilaton fluid}) winning over the dilaton-axion coupling, and this causes $\chi$ to be driven to \emph{larger} values (so particle masses decrease). Because the existence of the excursion depends only on the change in tracker solutions as one moves from radiation to matter domination, it does not go away even in the limit $\zeta \to 0$ \cite{Albrecht:2001xt}. Because of this there is a generic prediction for a nonzero mass difference between now and recombination and between now and nucleosynthesis, whose size is set by the coupling $\bfg$. For $\mathbf{g} = -10^{-3}$ and $\zeta = 0$ the mass shift is $(m_{\rm now} - m_{\rm BBN})/m_{\rm BBN}=-0.003\%$. Except for a window around $\zeta \sim 0.01$ the mass difference at nucleosynthesis is expected to be larger than it is at recombination.

We see that the small size for $\bfg$ dictated by solar-system tests makes the variation in particle masses in cosmology quite small. The same need not be true for Dark Matter, since the $\chi$-dependence of its mass is controlled by the size of the dilaton-axion coupling parameter $\zeta$ (and not $\bfg$). (See Figure \ref{fig:No well W and axion mass} for a plot of how $W(\chi)$ and $\mfm$ evolve.) The axion-dilaton coupling begins to be visible in the angular and matter power spectrum when $\Delta m / m_o \sim \mathcal{O} (0.01\%)$. This differs somewhat from what was found in \cite{Smith:2024ayu} because we here have no direct axion-baryon interactions and explore weaker dilaton-baryon interactions. 

Although too small to be measured now, the robustness of the prediction of mass excursions makes it more interesting to develop observational ways to confirm or deny their presence. They are likely to be much larger in models like those explored in \cite{Burgess:2021obw, Smith:2024ayu}, where $\bfg$ is much larger, in which case some sort of mechanism is needed to screen the dilaton's interactions on solar system scales. One such a mechanism \cite{Brax:2023qyp} introduces axion-matter couplings so that the axion-dilaton interactions act to screen matter-dilaton couplings for macroscopic objects. In this case the presence of axion-matter interactions likely combines with larger values of $\bfg$ to give more easily probed effects, though a definitive prediction requires fully incorporating the screening mechanism to determine how much it changes what is expected for cosmology.

\begin{figure}[hbtp!]
    \begin{subfigure}[b]{0.99\textwidth}
        \centering
    \includegraphics[width = \linewidth]{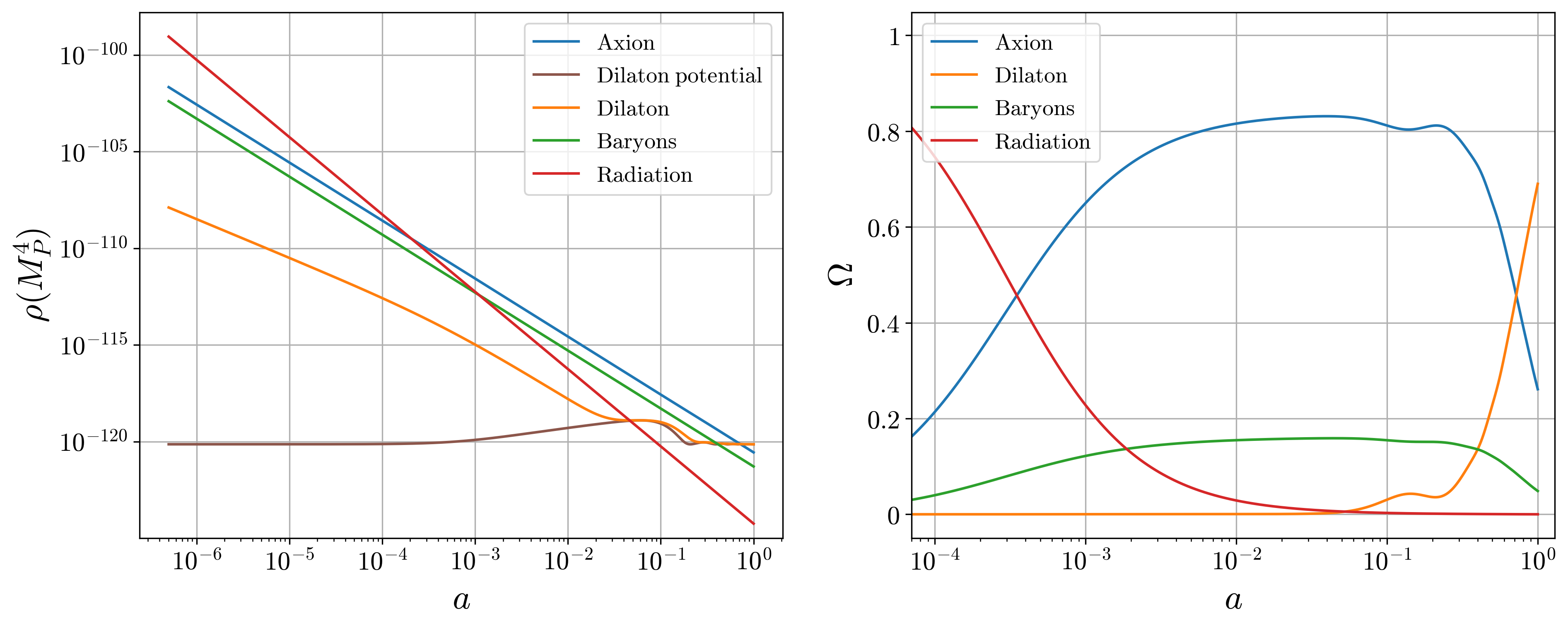}
    \end{subfigure}
    \hfill
    \begin{subfigure}[b]{0.99\textwidth}
        \centering
    \includegraphics[width = \linewidth]{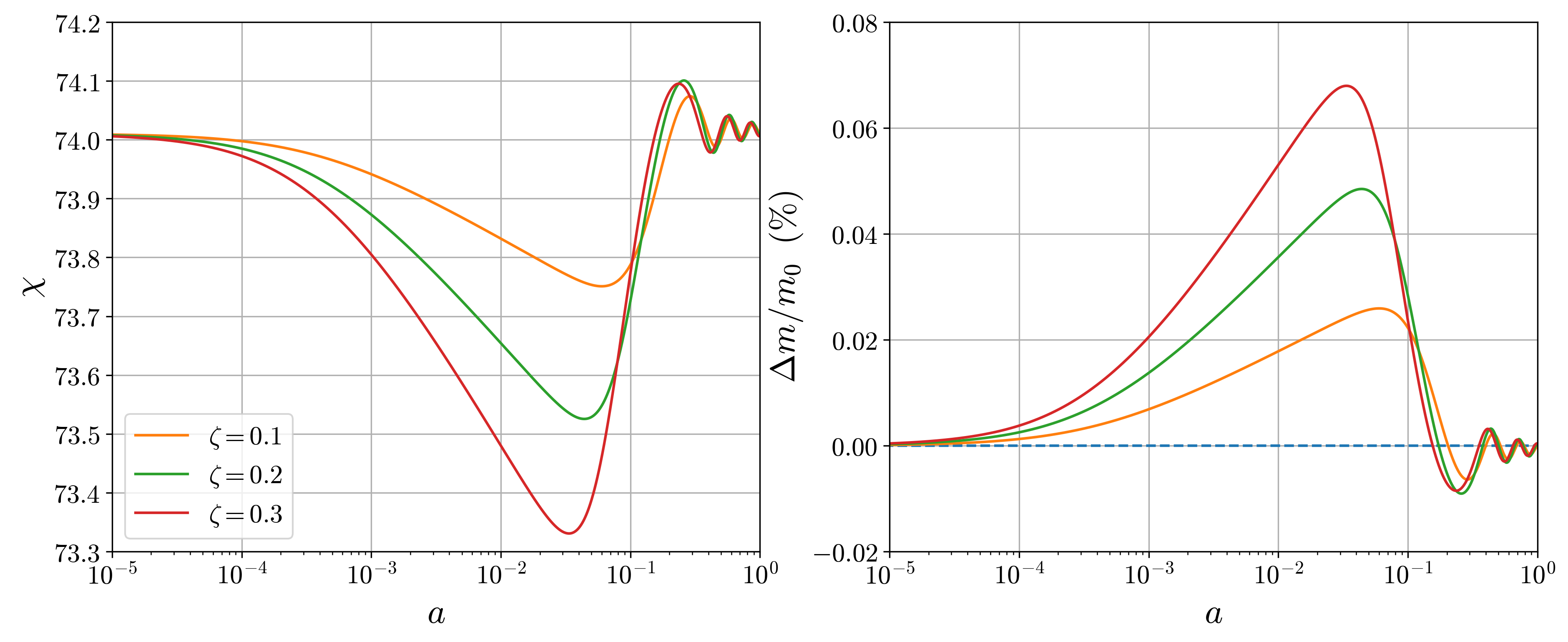}
    \end{subfigure}
    \hfill
    \begin{subfigure}[b]{0.99\textwidth}
        \centering
    \includegraphics[width = \linewidth]{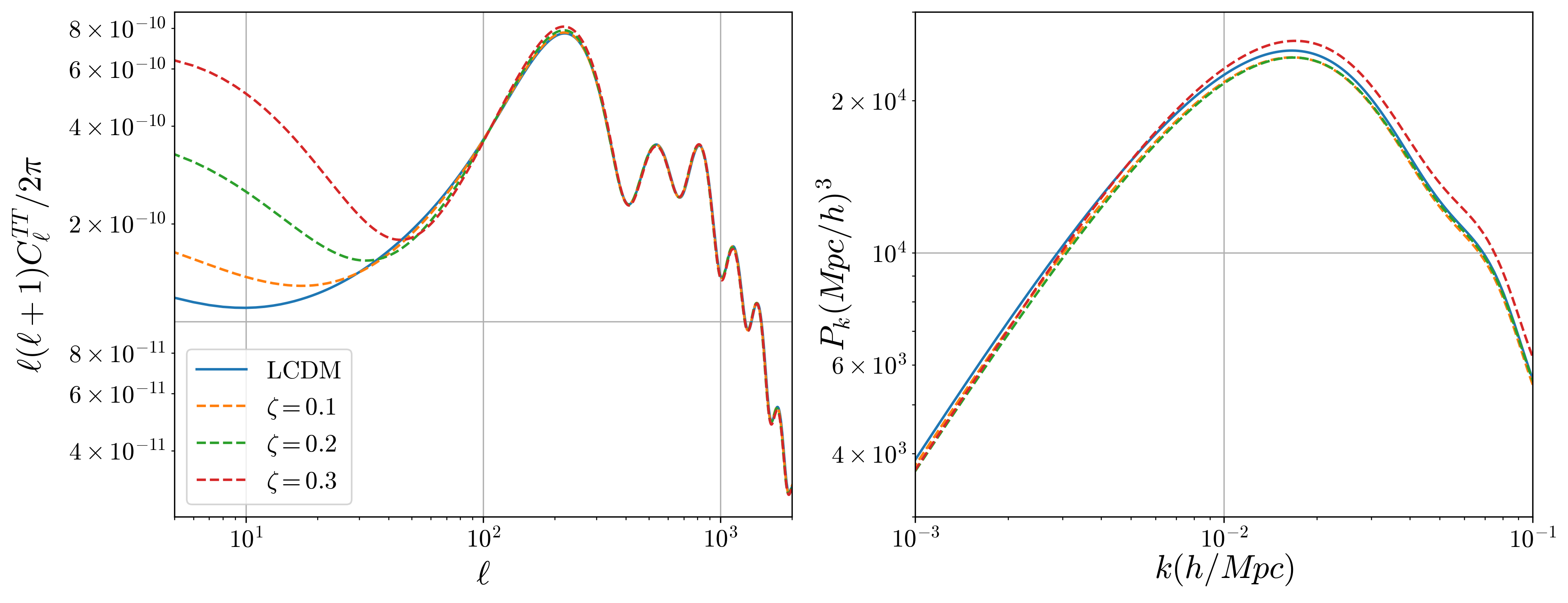}
    \end{subfigure}
    \hfill
    \caption{Background and perturbative level plots of the axio-dilaton cosmology with the inclusion of the dilaton potential well. Top row shows the evolution of the background energy densities when $\zeta = 0.1$. Middle row shows the evolution of the dilaton field and associated baryon masses for a range of $\zeta$. Bottom row shows the corresponding angular and matter power spectra with the $\Lambda\rm CDM$ best-fit in solid blue. In all cases $\mathbf{g}=-10^{-3}$. }
    \label{fig:Well}
\end{figure}

Figure \ref{fig:Well} shows how the above predictions change once $u_1$ and $u_2$ are nonzero so that the dilaton potential has a local minimum, as in \pref{YogaPot} and \pref{Uquad}. 
%
%\begin{align}
%    V(\chi) = Ue^{-4\zeta\chi} \qq{and} U = V_0\left[1-u_1\chi+\frac{u_2}{2}\chi^2\right].
%\end{align}
%
Because we explore smaller values of $\zeta$ than used in Yoga models we choose the coefficients $u_i$ to ensure that the value of $\chi$ at the minimum is somewhat larger, $\chi_{\min}\sim 70$, so that the value of $\tau = e^{\zeta \chi_{\rm min}}$ is again of order $\MPL^2/\MEW^2$. The different curves again correspond to a range of choices for $\zeta$, although the values chosen here run through a range larger than those that were considered in fig,~\ref{fig:No well}. 

The dilaton is again started off close to $\chi_{\rm min}$ since this also becomes its present-day value. It is given an initial kinetic energy that is much larger than its potential energy, but small enough that it remains less than a few percent of the energy budget at nucleosynthesis, and arranged to ensure the dilaton eventually gets caught in the potential's local minimum at late times. (This is easier than one might think for the same reasons as were found in \cite{Burgess:2021obw}: both because of Hubble friction and because the axion-generated potential easily dominates in the early universe and pushes in the opposite direction than $V(\chi)$ and the matter-generated dilaton potential.) 

The first row of plots again show the density of the various components of the cosmic fluid, again showing how the energy density in the axions falls like $1/a^3$ as required for Dark Matter. In this version of the cosmology the energy densities again largely follow scaling solutions, though this changes for the dilaton once the potential's non-exponential form begins to dominate the axion-generated and matter-generated potentials and $\chi$ starts to approach the minimum. 

In the example shown these oscillations are never completely damped out, making tests of correlated oscillatory time-dependence of masses (such as from quasar spectra) and the Dark Energy density particularly interesting.

The second row of plots again show how $\chi$ changes over time and the evolution of ordinary particle masses to which it gives rise. A dilaton excursion again generically arises after radiation-matter equality, for the same reasons as were described earlier. For $\zeta = 0.1$ (the only value plotted in both fig.~\ref{fig:No well} and fig.~\ref{fig:Well}) the dilaton excursion is about half as large as what was found when the potential was purely exponential.\footnote{Having the local minimum in the potential means we are no longer free to shift $\chi$ to make its current value zero.} The general trend towards smaller values of $\chi$ near recombination -- and so towards larger particle masses -- is again present, for the same reasons as before (for larger $\zeta$ the axion-generated potential dominates and pushes $\chi$ to smaller values). The main new effect is the appearance of late-time damped oscillations in $\chi$ (and so also in particle masses) as it seeks the potential's minimum.

The implications for the CMB and the matter power spectrum are shown in the figure's bottom row. The effect of having a local minimum in the dilaton potential is to stabilize the dilaton's late-time dynamics and this reduces the effects of its motion on the CMB. (This is why we show results using larger values for $\zeta$.) This is seen in the bottom row of fig.~\ref{fig:Well} where the deviations from the $\Lambda$CDM best-fit are reduced for the same $\zeta$ compared to those in fig.~\ref{fig:No well}. 

\begin{figure}[hbtp!]
    \begin{subfigure}[b]{0.99\textwidth}
        \centering
    \includegraphics[width = \linewidth]{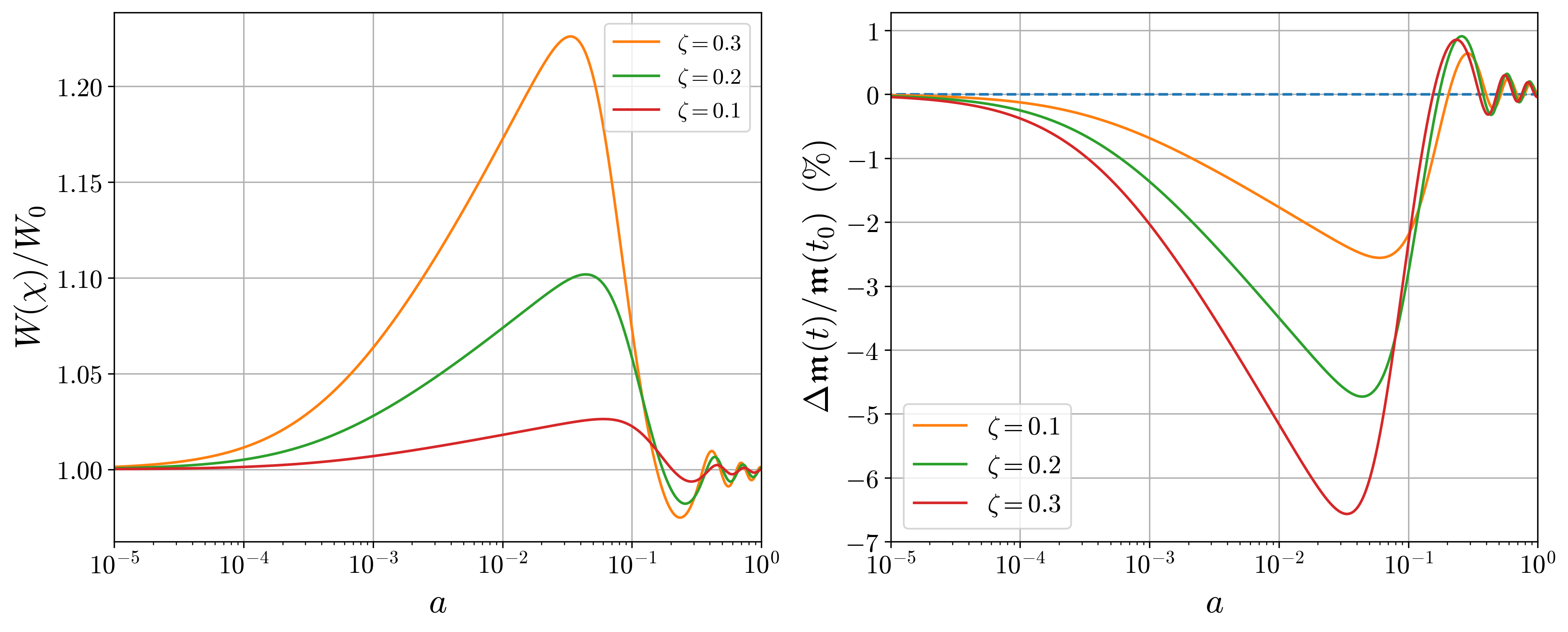}
    \end{subfigure}
    \hfill
    \caption{Fractional changes in the axion decay constant $W(\chi)$ and effective mass $\mfm(t)$ with scale factor when the dilaton is trapped in a local minimum of its potential described by (\ref{YogaPot}) and (\ref{Uquad}).}
    \label{fig:well W and axion mass}
\end{figure}

Fig.~\ref{fig:well W and axion mass} shows the evolution of the decay constant and the axion mass in this case. In both cases the axion-dilaton coupling works to decrease the value of the dilaton field. When this effect dominates over the baryon-dilaton coupling (which acts in the opposite direction) this causes a decrease in the effective axion mass, $\mfm(t) = m_\mfa e^{\zeta\chi}$, and hence energy density of the axion fluid. This is suggestive given the observed preference in DESI BAO data for energy-momentum flow from CDM to dark energy in \cite{Giare:2024smz}, and further motivates exploring axion-dilaton couplings that are stronger than the baryon-dilaton coupling. A full data analysis for the specific CDM-dark energy interactions in an axio-dilaton framework would be warranted to see if the same preference is observed.

\subsection{Structure growth}
 
\begin{figure}[hbtp!]
    \begin{subfigure}[b]{0.99\textwidth}
        \centering
    \includegraphics[width = \linewidth]{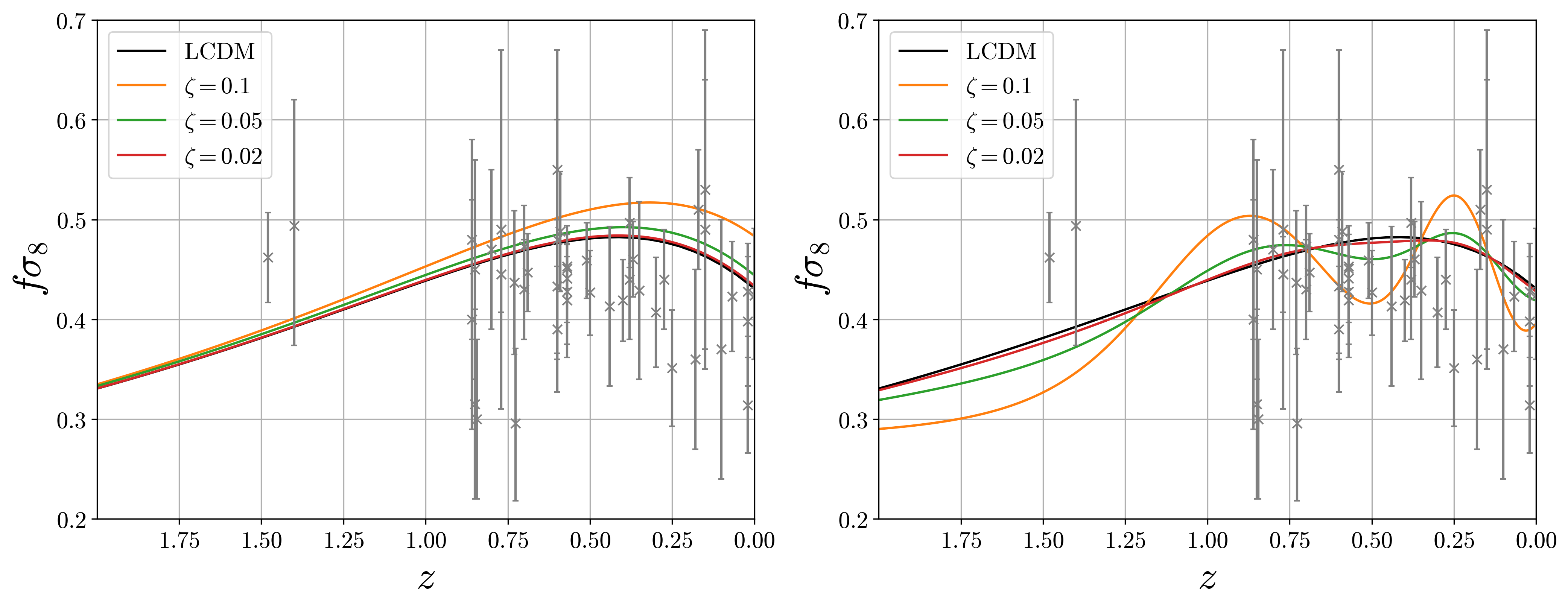}
    \end{subfigure}
    
    \caption{Left panel shows the structure growth parameter $f\sigma_8$ for the pure exponential potential and the right panel shows the same when the dilaton is trapped in a quadratic minimum. The $\Lambda$CDM best-fit is shown in black. The data points in grey are taken from \cite{Marulli:2020uyy}.}
    \label{fig:fsigma8}
\end{figure}

We have also used CLASS to compute the evolution of late-time structure. Figure \ref{fig:fsigma8} shows the evolution predicted by these equations for the parameter $f\sigma_8$, defined by 
\begin{equation}
    f\sigma_8 = \frac{\sigma_8(z,k_{\sigma8})}{\mathcal{H}}\frac{\delta_m'(z,k_{\sigma8})}{\delta_m(z,k_{\sigma8})} \, ,
\end{equation}
 where $k_{\sigma8} = 0.125h~$Mpc$^{-1}$ and $\delta_m = (\delta \rho_\ssB +W^2\delta \rho_\mfa)/(\Bar{\rho}_B+W^2\Bar{\rho}_\mfa)$. This variable provides an observable measure of clustering, for which a range of measurements are given as points with largish error bars. The curves in the left-hand panel give the prediction for a purely exponential dilaton potential while those in the right-hand panel give the result when the dilaton potential has a local minimum. The various curves correspond to different choices for $\zeta$, with the $\Lambda$CDM case also shown for comparison. 
 
 These figures show that increasing the dilaton-axion coupling $\zeta$ enhances structure growth, for the reasons described earlier: the additional attractive force in addition to gravity. The case where the dilaton potential has a minimum has more complicated behaviour, due to the background effect caused by the coupling of the axion to the oscillations of the dilaton within its potential well between recombination and today. The amplitude of these dilaton oscillations increase with increased $\zeta$ independent of the size of the correlated oscillations in particle masses. The right panel shows how this leads to larger and larger oscillations in the combined growth rates between species. In particular the prediction falls above or below the $\Lambda$CDM value at different times, and unlike for the exponential case can be smaller than $\Lambda$CDM at present. For instance, when $\zeta = 0.02$, $\sigma_8(z = 0)=0.824$ while when $\zeta = 0.1$, $\sigma_8(z = 0)=0.815$ showing a net decrease in the amplitude of structure growth today. 
 
 It is interesting to compare the modifications in structure growth for the case when the dilaton is trapped within its quadratic minimum with the effects on the CMB angular power spectrum in the bottom row of figure \ref{fig:Well}. The case $\zeta = 0.1$ is enough to give large deviations in structure growth while producing minimal deviations from $\Lambda$CDM for the low $\ell$ regime in the angular power spectrum. A full data analysis to examine whether these deviations could fit multiple combinations of cosmological datasets is therefore of interest for these models. Future experiments sensitive to the integrated Sachs-Wolfe effect such as CMB cluster cross-correlations are predicted to be able to place tighter constraints on these deviations, as discussed in \cite{Ballardini:2017xnt, Carron:2022eum} .

\section{Summary and Conclusions}
\label{sec:Conclusions}
To summarize, in this paper we proposed that the two--field axion-dilaton system provides a cosmologically viable model for the dark sector, in which the axion field plays the role of Dark Matter and the dilaton field is the Dark Energy. The resulting interacting Dark Energy model has the following ingredients, some of which distinguish it from the interacting Dark Energy models usually studied in the literature:

\begin{itemize}
    \item Dark Energy couples to Dark Matter. This coupling is determined by the function $W(\chi)$, and for $W(\chi) \propto e^{-\zeta \chi}$, the interaction strength is determined by the parameter $\zeta$. This interaction causes the (effective) mass of the axion Dark Matter field to vary as the dilaton evolves. 

    \item Dark Energy couples to baryons. The interaction strength is determined by $\bfg$. As a consequence, the masses of baryons vary as the dilaton evolves. The coupling of the dilaton to baryons imply a long--range fifth force between baryons, and we choose this coupling small enough to have escaped notice in solar system tests of GR. In the scenario studied here, the axion does not couple to baryons. 

    \item Typically, the potential for the dilaton is of exponential form, but breaking of scale--invariance allows the potential to obtain a (local) minimum. We have studied both cases in this paper: the pure exponential case and the Yoga example, in which the exponential is multiplied by a prefactor which depends on the dilaton, allowing the potential to have a minimum at finite value for $\chi$. The evolution of the dilaton, and hence particle masses, is different in these two scenarios. 

\end{itemize}
    
As mentioned above, the dilaton is light enough to mediate a long-range force and so the parameter $\bfg$ must be small enough to escape notice in solar-system experiments ($\abs{\mathbf{g}}\lesssim 2\times10^{-3}$). We have chosen in our analysis $\bfg = -10^{-3}$. The variation of the baryons masses depend then on the dilaton evolution. In the case of a pure exponential potential we find that for $\zeta \leq 0.1$, the baryon masses vary less than 0.04 percent from deep inside the radiation dominated epoch until today (see Fig. \ref{fig:No well}). The presence of a local minimum in the potential for the dilaton does not stop the dilaton from going through a transient excursion after radiation-matter equality, due to the transition between early-universe tracker solutions, but introduces oscillations in $\chi$ (and particle masses) --  albeit with a small amplitude -- at late times as it returns to the potential's minimum (see Fig. \ref{fig:Well}). 

Because of the couplings to matter, we find that the largest effect on the CMB anisotropy power spectrum is at low multipoles. The variation of baryon and Dark Matter masses causes the gravitational potentials to evolve even in the matter dominated epoch according to the modified Einstein equations (\ref{Perturbed Friedmann2}), (\ref{perturbed 0-i2}) and (\ref{perturbed i-j2}), causing a large contribution to the integrated Sachs--Wolfe effect. Similar studies on the integrated Sachs-Wolfe effect in the presence of modified gravity have been performed in for example \cite{Chudaykin:2025gdn}. These studies show it is complicated to draw conclusions from these types of deviations when comparing with multiple cosmological datasets. In order to infer if these deviations from $\Lambda$CDM can  describe the cosmology better, a full data analysis would be required, which we leave for future work.

We also calculated the quantity $f\sigma_8$, where $f$ is the growth rate of density perturbations in matter and $\sigma_8$ the normalization of the matter power spectrum at $R=8 h^{-1}$Mpc, as a function of redshift. We find that for a pure exponential potential, this quantity behaves qualitatively similarly as in the $\Lambda$CDM model, but is usually larger. This behaviour is similar to what has been found in other interacting dark energy models \cite{Zhumabek:2023ejd,vandeBruck:2022xbk,Pooya:2024wsq}. If the potential has a local minimum, the dilaton oscillates around this minimum. As a consequence, both the background evolution and the growth rate evolution is modified, resulting in oscillatory behaviour of the quantity $f\sigma_8$, for which the amplitude is determined by $\zeta$, as can be seen in Fig. \ref{fig:fsigma8}. This behaviour is significantly different from other interacting Dark Energy models. 

Although the dilaton is not screened in this paper, in a future publication \cite{screenedcosmo} we explore a model in which the dilaton has larger couplings to matter, but is screened to evade solar system tests. This allows mass-evolution and axion-oscillation effects to play larger roles during late time cosmology. 

\subsection*{Phantom Note Added}

We briefly comment here on the possibility of apparent phantom behaviour in models like this, in view of the DESI results \cite{DESI:2024mwx} that came out several months after this paper was posted. The main observation is that the dilaton-dependence we find here for the Dark Matter mass means that the actual equation of state parameter for the dilaton,  
\begin{equation}
 \omega_\chi(\chi) =\frac{\chi'^2-2a^2V(\chi)}{\chi'^2 + 2a^2V(\chi)} \geq -1\,, 
\end{equation}
can differ from the {\it effective} Dark Energy equation of state parameter $\omega_{\chi\,\text{eff}}$ one would infer if one were to mistakenly try to interpret the dark sector as a quintessence field coupled to ordinary Dark Matter (such as is done when using the $w_0w_a$DM model), along the lines discussed in e.g. \cite{Das:2005yj, Brax:2011qs, vandeBruck:2020fjo, Khoury:2025txd}. For the cases studied here the effective dark energy equation of state would be given by
\begin{equation}
    \omega_{\chi\,\text{eff}} = \frac{\omega_\chi(\chi)}{1+\left[e^{\zeta(\chi-\chi_0)} - 1\right]\frac{\rho_{\text{ax}0}}{a^3\rho_\chi}},
\end{equation}
where $\rho_{\text{ax}\,0} = Cm_\mfa/ W(\chi_0)$ and $\chi_0$ are respectively the dark matter energy density and value of the dilaton today.

\begin{figure}[hbtp!]
    \begin{subfigure}[b]{0.99\textwidth}
        \centering
    \includegraphics[width = \linewidth]{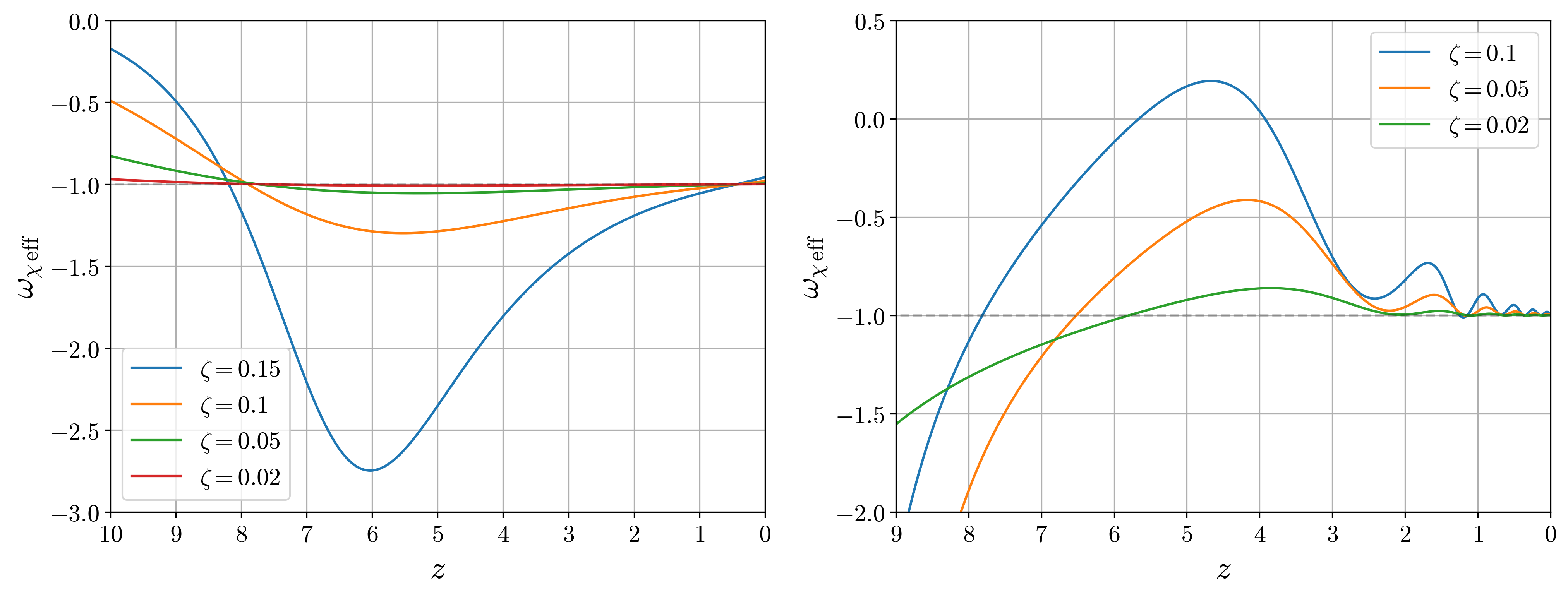}
    \end{subfigure}
    
    \caption{Effective dilaton equations of state as a function of redshift for the case of an exponential dilaton potential described by (\ref{YogaPot}) and (\ref{Uquad}) with $u_1 = u_2 = 0$ on the left and trapped in a local quadratic minimum described by (\ref{YogaPot}) and (\ref{Uquad}) on the right.}
    \label{fig:effective eos}
\end{figure}

Figure \ref{fig:effective eos} plots this effective equation of state vs redshift for the two classes of models considered here. These plots show how $\omega_{\chi\,{\rm eff}}$ can easily take values less than -1 (the `phantom' regime) despite $\omega_\chi \geq -1$ always being true. In particular $\omega_{\chi \text{eff}} < -1$ can hold right up to relatively small redshifts (around $z\sim 0.5$) in the case with a pure exponential dilaton potential for all coupling strengths, remaining in the phantom regime well into matter domination. When the dilaton has a local potential minimum into which to be trapped, there are much smaller crossings into apparently phantom territory as the dilaton oscillates during dark energy domination, but much deeper excursions at larger redshifts. The redshift at which this larger transition into the phantom regime occurs increases with dilaton-axion coupling strength.

\section*{Acknowledgements}

We thank Elsa Teixeira for the useful numerical resources. This work evolved out of discussions at the Astroparticle Symposium at the Institut Pascal. CB, CvdB and ACD thank the Institut Pascal for their hospitality during the programme. MM is also grateful for the hospitality of Perimeter Institute where part of this work was carried out. AS is supported by the W.D. Collins Scholarship. CvdB is supported by the Lancaster–Sheffield Consortium for Fundamental Physics under STFC grant: ST/X000621/1. ACD is partially supported by the Science and Technology Facilities Council (STFC) through the STFC consolidated grant ST/T000694/1. CB's research was partially supported by funds from the Natural Sciences and Engineering Research Council (NSERC) of Canada. MM is supported Kavli IPMU which was established by the World Premier International Research Center Initiative (WPI), MEXT, Japan. This work was also supported by a grant from the Simons Foundation (1034867, Dittrich). Research at the Perimeter Institute is supported in part by the Government of Canada through NSERC and by the Province of Ontario through MRI. 

\appendix

\bibliographystyle{JHEP}
\bibliography{bibliography}

\end{document}